\documentclass[english,preprint,superscriptaddress,showpacs,prb]{revtex4}%
\usepackage[T1]{fontenc}
\usepackage[latin1]{inputenc}
\usepackage{amsmath}
\usepackage{graphicx}
\usepackage{amssymb}
\usepackage{amsfonts}
\usepackage{babel}%
\providecommand{\U}[1]{\protect\rule{.1in}{.1in}}
\makeatletter
\makeatletter
\makeatletter
\makeatletter
\makeatletter
\makeatother
\makeatother
\makeatother
\makeatother
\makeatother
\begin{document}
\author{Alexey A. Kovalev}
\affiliation{Kavli Institute of NanoScience, Delft University of Technology, 2628 CJ Delft,
The Netherlands}
\author{Gerrit E. W. Bauer}
\affiliation{Kavli Institute of NanoScience, Delft University of Technology, 2628 CJ Delft,
The Netherlands}
\author{Arne Brataas}
\affiliation{Department of Physics, Norwegian University of Science and Technology, NO-7491
Trondheim, Norway }
\affiliation{Center for Advanced Study at the Norwegian Academy of Science and Letters,
Drammensveien 78, NO-0271 Oslo, Norway}
\title{Current-driven ferromagnetic resonance, mechanical torques and rotary motion
in magnetic nanostructures}

\begin{abstract}
We study theoretically the detection and possible utilization of electric
current-induced mechanical torques in ferromagnetic-normal metal
heterostructures generated by spin-flip scattering or the absorption of
transverse spin currents by a ferromagnet. To this end, we analyze the DC
voltage signals over a spin valve driven by an AC current. In agreement with
recent studies, this {}``rectification'', measured as a function of AC
frequency and applied magnetic field, contains important information on the
magnetostatics and --dynamics. Subsequently, we show that the vibrations
excited by spin-transfer to the lattice can be detected as a splitting of the
DC voltage resonance. Finally, we propose a concept for a spin-transfer-driven
electric nanomotor based on integrating metallic nanowires with carbon
nanotubes, in which the current-induced torques generate a rotary motion

\end{abstract}
\date{\today{}}

\maketitle

\section{Introduction}

An electromotor is an apparatus that generates rotational motion with
electrical currents, as demonstrated by Faraday in 1821. Under
miniaturization, the torques generated by Ørsted magnetic fields scale
unfavorably compared to the increased friction, rendering the Faraday motor an
unfeasible concept at the nanoscale. Alternative concept for nanomachines are
based on electrostatic forces,\cite{Fennimore:nat03} thermal
fluctuations,\cite{Xu:cm06}, torques induced by circularly polarized
light\cite{Kral:prb02} and angular momentum transfer by spin-polarized
currents. \cite{Mohanty:prb04,Malshukov:prl05,Kovalev:jjap06,Stiles:jap02} In
this article, we elaborate on the last idea.

A spin-polarized current carries an angular momentum current:
\begin{equation}
T_{SP}=PI\hbar/(2e)=PT_{0} \label{tau0}%
\end{equation}
where $P=\left(  I_{\uparrow}-I_{\downarrow}\right)  /I$ is the polarization
of the charge current $I=I_{\uparrow}+I_{\downarrow}$, $e$ is the electron
charge and $T_{0}=I\hbar/(2e)$. If transferred completely to the lattice,
mechanical torques are created of the same magnitude, which can be relatively
large in nanoscale structures. Since spin currents are routinely excited in
magnetoelectronic devices such as spin valves, we pose here the question
whether current-induced mechanical torques can be detected and utilized in
such structures. We conclude that the resonant magnetomechanical coupling
studied earlier\cite{Kovalev:apl03,Kovalev:jjap06} should indeed be observable
in spin valve structures, paving the way for applications such as high
frequency actuators and transducers of mechanical motion. Furthermore, we
propose a design for a spin-transfer driven electric nanomotor based on carbon
nanotubes.\cite{Kral:prb02}

The resonant rectification of a current in spin valve as a function of an
applied AC frequency has been found experimentally to form a rich source of
information about the magnetization dynamics in spin valve
structures.\cite{Tulapurkar:nat05,Sankey:prl06} Kupferschmidt \textit{et al}.
found theoretically that the spin-pumping by the magnetization
dynamics\cite{Tserkovnyak:prl02} significantly modify these
spectra.\cite{Kupferschmidt:2006} We suggest that the AC-DC conversion in spin
valves can be used to detect vibrational modes excited by the spin-polarized currents.

In this manuscript, we address the theory of spin valves excited by AC
currents, show how to include the effects of the magnetovibrational coupling,
and predict signatures of the current-induced mechanical torques. We also
share our ideas how these torques could drive a rotary (rather than
vibrational) motion, \textit{i.e.} an electric nanomotor. The manuscript is
organized as follows. In Sec. II, we calculate (position-dependent) mechanical
torques generated by the spin-flip dissipation of a spin current injected by a
ferromagnet into a normal metal. In Sec. III, we study the spin-transfer
mechanical torques resulting from the absorption of transverse spin-currents
by a ferromagnet. We suggest to employ the recently reported diode
effect\cite{Tulapurkar:nat05,Sankey:prl06} in F(erromagnet)%
$\vert$%
N(ormal)%
$\vert$%
F(erromagnet) metal spin valves to detect the vibrations created by
spin-transfer mechanical torques. We calculate the nonlinear DC voltage
emanating from the spin-transfer driven ferromagnetic resonance with and
without the resonant magnetovibrational coupling. Spin transfer torques
non-collinear with planes formed by principal axis of anisotropies can deform
the resonant line shape of DC voltage as a function of the AC current bias
frequency. In Appendix B and C, we calculate the extra DC voltage caused by
spin-pumping and prove that it can play a key role in sufficiently thin films.
In Appendix B, we also conclude that the magnetovibrational coupling is
observable by virtue of the spin-pumping even in asymmetric N%
$\vert$%
F%
$\vert$%
N heterostructures. In Sec. IV, we propose a spin-transfer driven nanomotor
concept based on integrating metallic nanowires with carbon nanotubes.

\section{Mechanical torques due to dissipation of spin currents}

\begin{figure}[ptb]
\centerline{\includegraphics[scale=0.7]{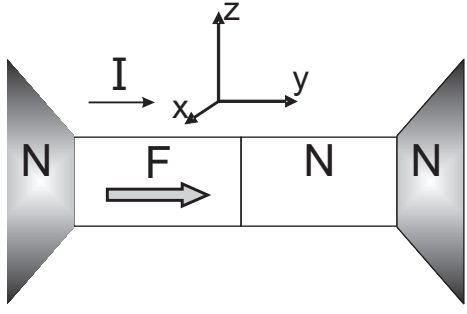}}\caption{Spin transfer from
spin-polarized currents to the lattice in a heterostructure consisting of two
(ferromagnetic and normal metal) sections connected to mechanically clamped
normal and ferromagnetic metal reservoirs. Polarized currents lead to torques
twisting the middle section.}%
\label{rev}%
\end{figure}

Consider a normal-metal diffusive wire or nanostructured pillar into which a
spin accumulation has been injected via an electrically biased ferromagnetic
contact (Fig. \ref{rev}). Spin $\mathbf{I}_{s}$ and charge $I_{0}$ currents
can be conveniently related via $2\times2$-matrices in Pauli spin space
$\widehat{I}=(\widehat{1}I_{0}+\widehat{\mathbf{\sigma}}\cdot\mathbf{I}%
_{s})/2,$ where $\widehat{\mathbf{\sigma}}$ is the vector or Pauli spin
matrices and $\widehat{1}$ is the $2\times2$ unit matrix. Spin-orbit
interactions or magnetic impurities cause spin-flip scattering that can be
parametrized by a spin-flip relaxation time $t_{sf}$. $\boldsymbol{\mu}_{s}$,
the local (vector) spin accumulation is related to the spin current density
$\mathbf{j}_{s}=\mathbf{I}_{s}/(eS)$ ($e$ is the electron charge and $S$ is
the cross-section of the wire) by the angular momentum conservation law:
\begin{equation}
\frac{\partial}{\partial t}\boldsymbol{\mu}_{s}+\frac{\partial}{\partial
y}\frac{\mathbf{j}_{s}}{\mathcal{N}}=\frac{\boldsymbol{\mu}_{s}}{t_{sf}}.
\label{diff0}%
\end{equation}
The dissipated angular momentum per unit length, $\boldsymbol{\mathbf{\tau}%
}=\left(  \hbar/2\right)  S\mathcal{N}\boldsymbol{\mu}_{s}/t_{sf},$ where
$\mathcal{N}$ is the density of states at the Fermi-level, is transferred as a
mechanical torque to the lattice. In the configuration sketched in Fig.
\ref{rev}, the injected spin accumulation $\left\vert \mathbf{\mu}%
_{s}\right\vert =\mu_{s}$ and the mechanical torque are polarized in the
\textit{y}-direction. Newton's Law for the mechanical motion of the substrate
then reads\cite{Landau:59}
\begin{equation}
\rho I\frac{\partial^{2}\varphi(y,t)}{\partial t^{2}}=C\frac{\partial
^{2}\varphi(y,t)}{\partial y^{2}}+\frac{\hbar}{2}S\mathcal{N}\frac{\mu
_{s}(y,t)}{t_{sf}} \label{mech0}%
\end{equation}
where $\varphi(y,t)$ is the angle of torsion, $I=I_{x}+I_{z}$ $\left(
I_{z}=\int x^{2}dzdx,I_{z}=\int z^{2}dzdx\right)  $ is the moment of inertia
of the cross-section at $y$ relative to its center of mass, $\rho$ is the mass
density, $C$ is an elastic constant defined by the shape and material of the
wire ($C=\mu R^{4}/2$ for a circular cross section with radius $R$, $\mu$ is
the Lam$\acute{\mathrm{e}}$ constant). Eq. (\ref{mech0}) must be complemented
by the boundary conditions ($\varphi=0$) at the clamping points.

We now concentrate on a bimetal wire consisting of a ferromagnetic and a
normal metal (Fig. \ref{rev}). We assume for simplicity here that the bulk
resistances of the wires are much larger than the interface resistance (Ref.
\onlinecite{Kovalev:prb02} shows how interfaces can be taken into account) to
the extent that we may disregard the latter. The magnetization and mechanical
motion is much slower than the relaxation scattering, therefore, we can
consider only the parametrically stationary limit. The charge current density
$j_{0}$ is conserved ($\partial_{y}j_{0}=0$) and\ Eq. (\ref{diff0}) reduces
to
\begin{equation}
\frac{\partial}{\partial y}\mathbf{j}_{s}=\frac{\mathcal{N}\boldsymbol{\mu
}_{s}}{t_{sf}}=\frac{2\boldsymbol{\mathbf{\mathbf{\tau}}}}{\hbar S}\,.
\label{diff2}%
\end{equation}
In the normal metal, charge and spin currents are governed by Fick's Laws
$j^{N}=\mathcal{N}^{N}D^{N}\partial_{y}\mu_{0}^{N}$ and $\mathbf{j}_{s}%
^{N}=\mathcal{N}^{N}D^{N}\partial_{y}\boldsymbol{\mathbf{\mu}}_{s}^{N}$
respectively, where $D^{N}$ is the diffusion constant and the index $N$
indicates the normal metal, leading to the diffusion equations:%
\[
\frac{\partial^{2}}{\partial y^{2}}\mathcal{N}^{N}D^{N}\mu_{0}^{N}%
=0\,,\hspace{1cm}\frac{\partial^{2}}{\partial y^{2}}\mathcal{N}^{N}%
D^{N}\mathbf{\mu}_{s}^{N}=\mathcal{N}^{N}\mathbf{\mu}_{s}^{N}/t_{sf}^{N}%
=\frac{\mathbf{\tau}_{N}}{\frac{\hbar}{2}S}\,.
\]
where $\mathbf{\tau}_{N}$ is the mechanical torque per unit length for the
normal metal. In a ferromagnet ($F$), the particle and spin currents are
$j^{F}=(\mathcal{N}_{\uparrow}^{F}D_{\uparrow}^{F}\partial_{y}\mu_{\uparrow
}+\mathcal{N}_{\downarrow}^{F}D_{\downarrow}^{F}\partial_{y}\mu_{\downarrow
})/2$ and $\mathbf{j}_{s}^{F}=\mathbf{m}\partial_{y}(\mathcal{N}_{\uparrow
}^{F}D_{\uparrow}^{F}\mu_{\uparrow}-\mathcal{N}_{\downarrow}^{F}D_{\downarrow
}^{F}\mu_{\downarrow})/2$, where $D_{\uparrow(\downarrow)}^{F}$ is the
diffusion constants for spin-up (-down) electrons and $\mathcal{N}%
_{\uparrow(\downarrow)}^{F}$ is the corresponding spin-up (-down) density of
states. The diffusion equation in a ferromagnet then reads:
\[
\frac{\partial^{2}}{\partial y^{2}}(\mathcal{N}_{\uparrow}^{F}D_{\uparrow}%
^{F}\mu_{\uparrow}+\mathcal{N}_{\downarrow}^{F}D_{\downarrow}^{F}%
\mu_{\downarrow})=0\,,\hspace{1cm}\frac{\partial^{2}}{\partial y^{2}%
}(\mathcal{N}_{\uparrow}^{F}D_{\uparrow}^{F}\mu_{\uparrow}-\mathcal{N}%
_{\downarrow}^{F}D_{\downarrow}^{F}\mu_{\downarrow})=\mathcal{N}^{F}%
(\mu_{\uparrow}-\mu_{\downarrow})/t_{sf}^{F}\,,
\]
leading to%
\[
\frac{\partial^{2}}{\partial y^{2}}D^{F}(\mu_{\uparrow}-\mu_{\downarrow}%
)=(\mu_{\uparrow}-\mu_{\downarrow})/t_{sf}^{F}=\tau_{F}/\left[  \frac{\hbar
}{2}S\mathcal{N}^{F}\right]  \,,
\]
where $\mathbf{\tau}_{F}$ is the mechanical torques per unit length for the
ferromagnetic metal, $D^{F}=2D_{\uparrow}^{F}D_{\downarrow}^{F}\mathcal{N}%
_{F}/\left(  \mathcal{N}_{\uparrow}^{F}D_{\uparrow}^{F}+\mathcal{N}%
_{\downarrow}^{F}D_{\downarrow}^{F}\right)  $ and $\mathcal{N}^{F}%
=(\mathcal{N}_{\uparrow}^{F}+\mathcal{N}_{\downarrow}^{F})/2$.

\begin{figure}[ptb]
\centerline{\includegraphics{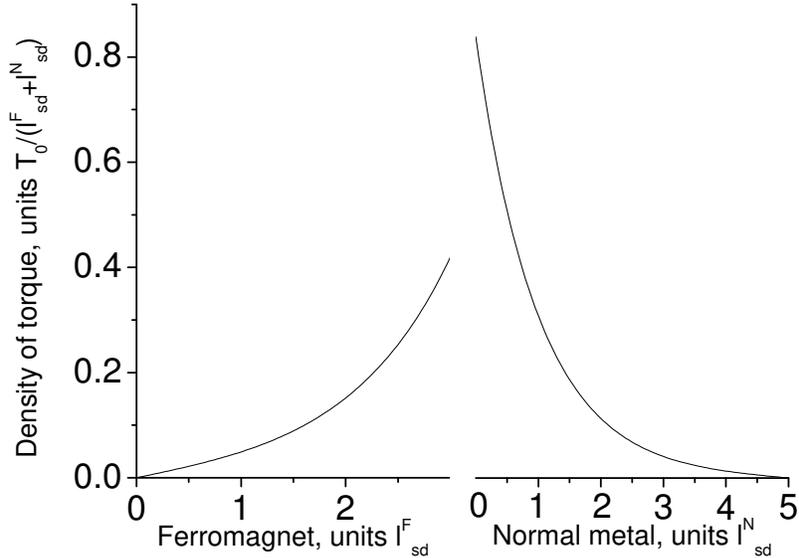}}\caption{The mechanical torque per unit
length along the F$|$N wire; $P=0.7$, $l_{sd}^{F}=10\text{nm}$, $l_{sd}%
^{N}=20\text{nm}$, $L_{F}/l_{sd}^{F}=3$, $L_{N}/l_{sd}^{N}=5$, and
$\upsilon=2$.}%
\label{wireT}%
\end{figure}

We can now calculate the mechanical torques by solving the diffusion
equations, requiring the continuity of distribution functions and conservation
of the spin and charge currents at the
interface.\cite{Brataas:epjb01,Kovalev:prb02} We also require no spin
accumulation at the connection with reservoirs. For the ferromagnetic ($y<0$)
and normal ($y>0$) metals we find, respectively:%
\begin{equation}
\tau_{F}\left(  y\right)  =T_{0}\frac{P\sinh\left[  (L_{F}+y)/l_{sd}%
^{F}\right]  /\sinh(L_{F}/l_{sd}^{F})}{l_{sd}^{F}\coth(L_{F}/l_{sd}%
^{F})+\upsilon l_{sd}^{N}\coth(L_{N}/l_{sd}^{N})}, \label{torqF}%
\end{equation}%
\begin{equation}
\tau_{N}\left(  y\right)  =T_{0}\frac{\upsilon P\sinh\left[  (L_{N}%
-y)/l_{sd}^{N}\right]  /\sinh(L_{N}/l_{sd}^{N})}{l_{sd}^{F}\coth(L_{F}%
/l_{sd}^{F})+\upsilon l_{sd}^{N}\coth(L_{N}/l_{sd}^{N})}, \label{torqN}%
\end{equation}
where $P=\left(  G_{\uparrow}-G_{\downarrow}\right)  /\left(  G_{\uparrow
}+G_{\downarrow}\right)  $ is the current polarization of the ferromagnet here
defined in terms of the spin-up and spin-down conductances $G_{\uparrow}%
\,\,$and $G_{\downarrow}$, $l_{sd}^{F}=\sqrt{D^{F}t_{sf}^{F}}$ and $l_{sd}%
^{N}=\sqrt{D^{N}t_{sf}^{N}}$ are the spin-diffusion lengths in the ferromagnet
and normal metals, respectively, and $T_{0}$ has been introduced in Eq.
(\ref{tau0}), $L_{F}$ and $L_{N}$ are the lengths of the ferromagnet and
normal metals respectively, and $\upsilon=\left[  \mathcal{N}_{N}t_{sf}%
^{F}\right]  /\left[  \mathcal{N}_{F}t_{sf}^{N}\right]  $. For the
configuration sketched in Fig. \ref{rev}, the mechanical torque is directed
along the $y$ axis. The torque density is discontinuous when $\upsilon\neq1$.
As shown in Fig. \ref{wireT}, the mechanical torques are enhanced close to the
F%
$\vert$%
N interface on the scale defined by the spin-diffusion length in both
ferromagnet and normal metals. The mechanical torques change sign with the
electric currents direction. 

Integrating Eqs. (\ref{torqF},\ref{torqN}) leads to the total torque acting on
the wire:%
\begin{equation}
T_{F}=T_{0}\frac{Pl_{sd}^{F}\tanh(L_{F}/2l_{sd}^{F})}{l_{sd}^{F}\coth
(L_{F}/l_{sd}^{F})+\upsilon l_{sd}^{N}\coth(L_{N}/l_{sd}^{N})}, \label{sumF}%
\end{equation}
\begin{equation}
T_{N}=T_{0}\frac{\upsilon Pl_{sd}^{N}\tanh(L_{N}/2l_{sd}^{N})}{l_{sd}^{F}%
\coth(L_{F}/l_{sd}^{F})+\upsilon l_{sd}^{N}\coth(L_{N}/l_{sd}^{N})}.
\label{sumN}%
\end{equation}
When $L_{F}\gg l_{sd}^{F}$ and $L_{N}\gg l_{sd}^{N},$ we obtain $T_{SP}%
=T_{F}+T_{N}=P\tau_{0}\,\,$as expected by the complete dissipation of the spin
current in this limit (Eq. (\ref{tau0})). By ultrasensitive displacement
detection, it should be possible to observe the mechanical strain caused by
the spin-flip torques\cite{Mohanty:prb04} in the setup of Fig. \ref{rev}.

\section{Generation and detection of the mechanical torques due to spin
transfer}

\begin{figure}[ptb]
\centerline{\includegraphics[scale=0.7]{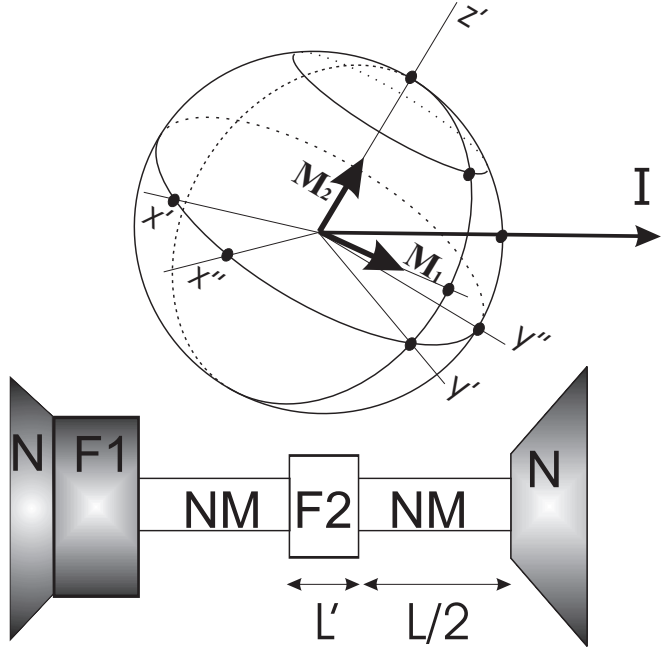}}\caption{Model sample to study
spin transfer to the lattice consisting of ferromagnets (F1 and F2) and NM,
\emph{i.e..} tunnel junctions or normal metal). F1 and the normal metal N are
electron reservoirs that are mechanically clamped. The vector diagram shows
the reference frames used in calculations with respect to the magnetizations.
The magnetization of F1 is fixed, $\mathbf{M}_{1}\equiv\mathbf{M}%
_{\mbox{fixed}},$ and that of F2 is variable, $\mathbf{M}_{2}\equiv\mathbf{M}%
$.}%
\label{rev1}%
\end{figure}

In the previous section, we studied mechanical torques arising from spin-flip
relaxation processes within the bulk of the metals. In contrast, in this
section, we will consider structures (see Fig. \ref{rev1}) in which the spin
transfer is dominated by dephasing processes (leading to absorption of
transversely polarized currents)\cite{Brataas:epjb01} at the normal
metal$|$ferromagnetic interfaces, whereas the spin-flip processes in the bulk
materials are disregarded (which can be repaired easily if necessary). Such a
device is superior to the wires of the previous section in generating
mechanical torques when the structures are smaller than the spin-diffusion
length. We propose here to induce and detect the magneto-vibrational modes
\cite{Kovalev:apl03} driven by spin-transfer torques in devices such as shown
in Fig. \ref{rev1}, \textit{i.e.} a doubly clamped heterostructure in the
shape of a bar with a ferromagnetic load in the center. In our set-up, the
mechanical torque is generated in the ferromagnet F2 by the transversely
polarized spin current (ferromagnet F1 is supposed to be clamped by the
substrate). The absorbed spin-angular momentum is transferred to the lattice
by the magnetic shape and crystal anisotropies. In the regime of the resonant
magnetovibrational coupling, the mechanical motion is strongly affected by the
magnetization dynamics created by spin-transfer torques,\cite{Kovalev:apl03}
which, in turn affect the transport properties. We predict that such
vibrations should be detectable by current-driven ferromagnetic resonance
experiments reported in Refs. \onlinecite{Tulapurkar:nat05} and
\onlinecite{Sankey:prl06}. We start by analyzing the DC voltage signals over a
spin valve resulting from the {}{}``rectification\textquotedblright\ of an AC
current by the precessing magnetization (without magnetovibrational coupling,
\textit{e.g.} setups from Refs. \onlinecite{Tulapurkar:nat05} and
\onlinecite{Sankey:prl06}) as an intermediate step. A general analysis of the
DC signals in the regime of the resonant magnetovibrational coupling is
subsequently given. We also estimate the maximum torques that can be created
in these structures.

The derivation below is carried out for a system in which the ferromagnet F2
is inserted into the normal metal island. However, the rectification of the AC
current and the excitation of magnetovibrational modes also occur (and can be
treated by our methods) in a system in which the ferromagnet is attached to
the side or on top of the normal metal wire ({}``spin-flip
transistor\textquotedblright).\cite{Wang,Wang:jjap06}

\subsection{Ferromagnetic resonance driven by spin-transfer torques}

\begin{figure}[ptb]
\centerline{\includegraphics{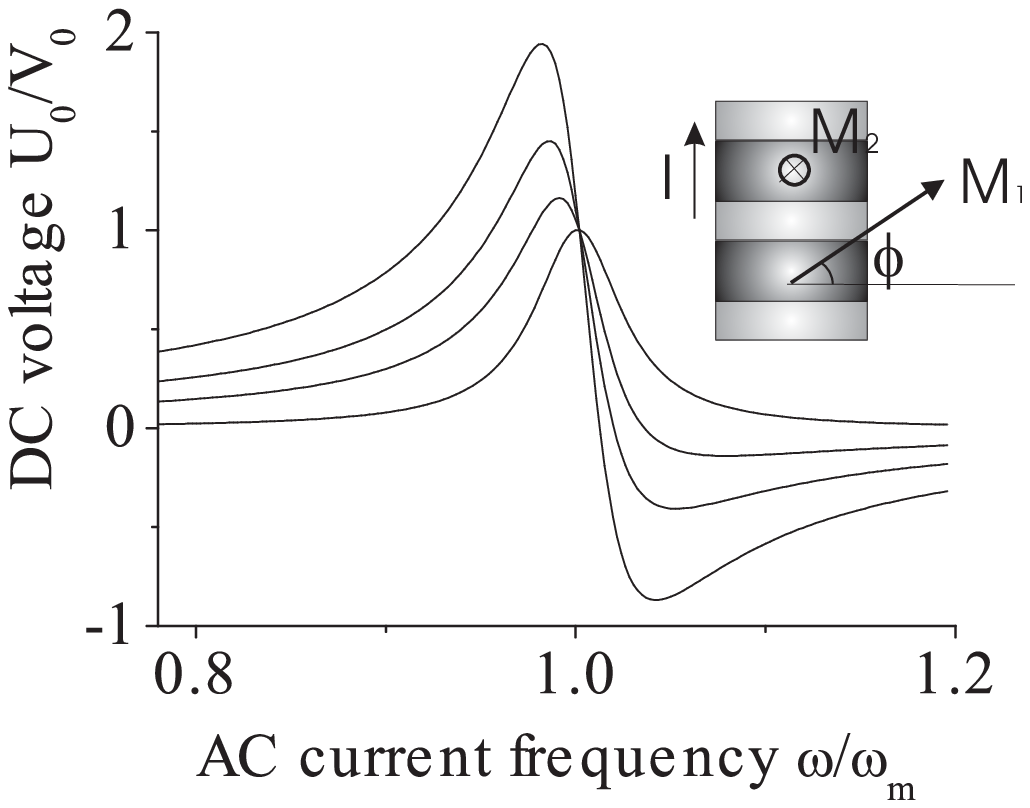}}\caption{The resonance peaks are skewed
when the spin-transfer torque does not lie in a plane formed by principal axes
of the shape anisotropy. We illustrate this by exciting the magnetization
$\mathbf{M}_{2}$ of the harder (resonant) layer with an easy plane anisotropy
by AC currents, while the soft layer magnetization is forced into the
directions $\phi=0:\pi/20:\pi/10:\pi/4$ . by an external magnetic field. Here
$\eta_{2}=0.$}%
\label{shape2}%
\end{figure}

Initially, we assume that the mechanical modes are not excited in the device
depicted in Fig. \ref{rev1}. One (the{}\textquotedblleft
hard\textquotedblright) layer magnetization of F1 is assumed completely fixed
by shape or crystal anisotropies to the direction $\mathbf{M}_{1}%
=\mathbf{M}_{\mathrm{fixed}}$ with unit vector $\mathbf{m}_{\mathrm{fixed}}$.
We are interested in the dynamics of the magnetization direction
$\mathbf{m=M}_{2}/M_{s}$ of the middle (\textquotedblleft
soft\textquotedblright) layer F2, where $\left\vert \mathbf{M}_{2}\right\vert
=M_{s}$ is the constant saturation magnetization, in the presence of an AC
electric current bias. We consider the soft layer to be at
resonance,\cite{Sankey:prl06} but note that the soft and hard layers can
exchange their roles as a function of frequency (\textit{e.g.} in Fig.
\ref{shape2}). We model the magnetization dynamics by the generalized
Landau-Lifshitz-Gilbert (LLG) equation in the macrospin approximation that is
augmented by the current-dependent spin-transfer
torque:\cite{Slonczewski:jmmm96}
\begin{equation}
\frac{d\mathbf{M}}{dt}=-\gamma\mathbf{M}\times\mathbf{H}_{\mathrm{eff}}%
+\frac{\alpha}{M_{\mathrm{s}}}\mathbf{M}\times\left(  \frac{d\mathbf{M}}%
{dt}\right)  +\gamma\frac{\hbar}{2e}\frac{I(t)}{V_{m}}\left[  \eta
_{1}\mathbf{m}\times(\mathbf{m}_{\mathrm{fixed}}\times\mathbf{m})+\eta
_{2}(\mathbf{m}_{\mathrm{fixed}}\times\mathbf{m})\right]  , \label{LLGapp}%
\end{equation}
where $I(t)$ is the time dependent current through the system, $\eta_{1}$
describes the efficiency of conventional spin transfer, and $\eta_{2}$
parametrizes an {}{}\textquotedblleft effective spin-transfer exchange
field\textquotedblright. When the spacer is an insulator, as in the experiment
of the Tsukuba group,\cite{Tulapurkar:nat05} the parameter $\eta_{1}$ is a
constant governed by the expressions such as derived by
Slonczewski.\cite{Slonczewski:prb05} The {}{}\textquotedblleft effective
field\textquotedblright\ $\eta_{2}$ has been less well investigated in
magnetic tunnel junctions and is treated here as an adjustable parameter. When
the spacer is a normal metal in a configuration sketched in Fig. \ref{rev1},
the LLG equation for the soft layer reads:
\begin{equation}
\frac{d\mathbf{M}}{dt}=-\gamma\mathbf{M}\times\mathbf{H}_{\mathrm{eff}}%
+\frac{\alpha}{M_{\text{s}}}\mathbf{M}\times\left(  \frac{d\mathbf{M}}%
{dt}\right)  -\frac{\gamma\hbar}{2eV_{m}}\mathbf{m}\times\left[
(\mathbf{I}_{s1}+\mathbf{I}_{s2})\times\mathbf{m}\right]  \label{LLGbody}%
\end{equation}
where the spin currents leaving the soft ferromagnet can be computed by
magnetoelectronic circuit theory\cite{Brataas:epjb01} as
\begin{equation}
I_{1(2)}=(G_{\uparrow}+G_{\downarrow})(\mu_{0}^{2(1)}-\mu_{0}^{1(2)}%
)+(G_{\uparrow}-G_{\downarrow})(\boldsymbol{\mathbf{\mu}}_{s}^{2(1)}%
-\boldsymbol{\mathbf{\mu}}_{s}^{1(2)})\cdot\mathbf{\mathbf{m}}
\label{chargebody}%
\end{equation}%
\begin{align}
\mathbf{I}_{s1(2)}  &  =\mathbf{\mathbf{m}}\left[  (G_{\uparrow}%
-G_{\downarrow})(\mu_{0}^{2(1)}-\mu_{0}^{1(2)})+(G_{\uparrow}+G_{\downarrow
})(\boldsymbol{\mathbf{\mu}}_{s}^{2(1)}-\boldsymbol{\mathbf{\mu}}_{s}%
^{1(2)})\right] \nonumber\\
&  -\left(  2\mathbf{\mathbf{m}}\times\boldsymbol{\mathbf{\mu}}_{s}%
^{1(2)}\right)  \times\mathbf{\mathbf{m}}G_{r}-\left(  2\mathbf{\mathbf{m}%
}\times\boldsymbol{\mathbf{\mu}}_{s}^{1(2)}\right)  G_{i}, \label{spinbody}%
\end{align}
where $\alpha$ is the Gilbert damping constant, $\gamma$ is the gyromagnetic
ratio, $G_{\uparrow}$ and $G_{\downarrow}$ describe the conventional
spin-dependent conductances limited by bulk and interface scattering and
$G_{\uparrow\downarrow}=G_{r}+iG_{i}$ is the interface mixing conductance of
the ferromagnet, $\mu_{0}$ and $\boldsymbol{\mathbf{\mu}}_{s}$ are the
chemical potential and spin-accumulation in the normal metals, respectively,
the spin $\mathbf{I}_{s1(2)}$ and charge $I_{1(2)}$ currents correspond to the
current (spin) flow into the normal metal node $1(2)$. It is shown in Appendix
A that Eqs. (\ref{LLGapp}) and (\ref{LLGbody}) are equivalent when we allow
for an angle dependence of the parameters $\eta_{1\left(  2\right)  }.$ The
efficiencies of the spin-transfer torque $\eta_{1}(\theta)$
\cite{Kovalev:prb02} and the {}{}\textquotedblleft effective spin-transfer
field\textquotedblright\ $\eta_{2}(\theta)$ mainly depend on the real and
imaginary part of the mixing conductance, respectively.

Choosing the $z^{\prime}$-axis along the equilibrium direction ($\mathbf{M}%
_{0}$) of the soft layer and the $x^{\prime}$-axis perpendicular to both
magnetizations (see Fig. \ref{rev1}), we expand the free energy close to the
equilibrium direction of the magnetization as%
\begin{equation}
F(\mathbf{M})=F(\mathbf{M}_{0})+N_{x^{\prime}}M_{x^{\prime}}^{2}%
/2+N_{y^{\prime}}M_{y^{\prime}}^{2}/2+N_{x^{\prime}y^{\prime}}M_{x^{\prime}%
}M_{y^{\prime}} \label{FREEapp}%
\end{equation}
such that the effective magnetic field
\begin{equation}
\mathbf{H}_{\mathrm{eff}}=-\partial F/\partial\mathbf{M=}-(N_{x^{\prime}%
}M_{x^{\prime}}+N_{x^{\prime}y^{\prime}}M_{y^{\prime}})\mathbf{x}^{\prime
}-(N_{y^{\prime}}M_{y^{\prime}}+N_{x^{\prime}y^{\prime}}M_{x^{\prime}%
})\mathbf{y}^{\prime} \label{Happ}%
\end{equation}
where $N_{x^{\prime}}$, $N_{y^{\prime}}$, and $N_{x^{\prime}y^{\prime}}$ are
parameters characterizing the energy of the macrospin excitations. The free
energy Eq. (\ref{FREEapp}) is not diagonal in the basis of the magnetizations
along the $x^{\prime}$ and $y^{\prime}$ axis, but can be diagonalized in
another basis rotated by the angle $\phi$ around the $z^{\prime}$%
\textit{-}axis, so that $N_{x^{\prime}}=N_{x}^{d}\cos^{2}\phi+N_{y}^{d}%
\sin^{2}\phi$, $N_{y^{\prime}}=N_{y}^{d}\cos^{2}\phi+N_{x}^{d}\sin^{2}\phi$,
$N_{x^{\prime}y^{\prime}}=(N_{x}^{d}-N_{y}^{d})\cos\phi\sin\phi$. Here,
$N_{x}^{d}$ and $N_{y}^{d}$ are the eigenvalues of the expansion tensor in Eq.
(\ref{FREEapp}). We consider here an arbitrary direction of an external
magnetic field $\mathbf{H}_{0}$. In general, the components of $\mathbf{H}%
_{0}$ contribute to Eq. (\ref{FREEapp}) in a non-trivial way. As an
illustration let us consider an external field along the $z$-axis, which can
be included as follows: $N_{x^{\prime}}=N_{x^{\prime}}^{0}+H_{0}/M_{s}$ and
$N_{y^{\prime}}=N_{y^{\prime}}^{0}+H_{0}/M_{s}$, where $N_{x(y)}^{0}$ are
elements of the magnetic anisotropy tensor.

The linear response of the magnetization to an AC current perturbation is
given by the response functions $\chi_{\mathrm{x^{\prime}I}}=(M_{x^{\prime}%
}/I)_{\omega}$ and $\chi_{\mathrm{y^{\prime}I}}=(M_{y^{\prime}}/I)_{\omega}$.
From Eqs. (\ref{LLGapp},\ref{Happ})
\begin{equation}
\chi_{\mathrm{x^{\prime}I}}(\omega)=\frac{\hbar\gamma\sin\theta}{2eV_{m}}%
\frac{i\eta_{2}\omega+\gamma M_{\mathrm{s}}\Gamma_{x^{\prime}}}{\omega
^{2}(1+\alpha^{2})-\omega_{\mathrm{m}}^{2}+2i\alpha^{\prime}\omega
\omega_{\mathrm{m}}}, \label{damping}%
\end{equation}%
\begin{equation}
\chi_{\mathrm{y^{\prime}I}}(\omega)=\frac{\hbar\gamma\sin\theta}{2eV_{m}}%
\frac{-i\eta_{1}\omega+\gamma M_{\mathrm{s}}\Gamma_{y^{\prime}}}{\omega
^{2}(1+\alpha^{2})-\omega_{\mathrm{m}}^{2}+2i\alpha^{\prime}\omega
\omega_{\mathrm{m}}}, \label{damping1}%
\end{equation}
where $V_{m}$ is the volume of the magnet, $\omega_{\mathrm{m}}^{2}=\gamma
^{2}N_{x}^{d}N_{y}^{d}M_{\mathrm{s}}^{2}=\gamma^{2}M_{\mathrm{s}}%
^{2}(N_{x^{\prime}}N_{y^{\prime}}-N_{x^{\prime}y^{\prime}}^{2})$,
$\Gamma_{x^{\prime}}=\eta_{1}(N_{y^{\prime}}+N_{x^{\prime}y^{\prime}}%
\alpha)+\eta_{2}(N_{x^{\prime}y^{\prime}}-N_{y^{\prime}}\alpha)$,
$\Gamma_{y^{\prime}}=\eta_{1}(N_{x^{\prime}y^{\prime}}+N_{x^{\prime}}%
\alpha)+\eta_{2}(N_{x^{\prime}}-N_{x^{\prime}y^{\prime}}\alpha)$, and the
damping parameter is modified by the anisotropies as
\begin{equation}
\alpha^{\prime}=\alpha\frac{(N_{x}^{d}+N_{y}^{d})/2}{\sqrt{N_{x}^{d}N_{y}^{d}%
}}=\alpha\frac{(N_{x^{\prime}}+N_{y^{\prime}})/2}{\sqrt{N_{x^{\prime}%
}N_{y^{\prime}}-N_{x^{\prime}y^{\prime}}^{2}}}. \label{renorm}%
\end{equation}
From Eqs. (\ref{damping},\ref{damping1}), we conclude that both the field
effect $\eta_{2}$, characteristic for tunnel junctions,\cite{Tulapurkar:nat05}
and non-collinear anisotropies $N_{x^{\prime}y^{\prime}}$ cause a phase shift
in the magnetization response $\chi_{\mathrm{y^{\prime}I}}(\omega)$ that is
relevant to the rectification effect, as demonstrated below. At resonance, the
parameters $\Gamma_{x^{\prime}}$ and $\Gamma_{y^{\prime}}$ are the components
of the out-of-phase magnetization response.

\begin{figure}[ptb]
\centerline{\includegraphics{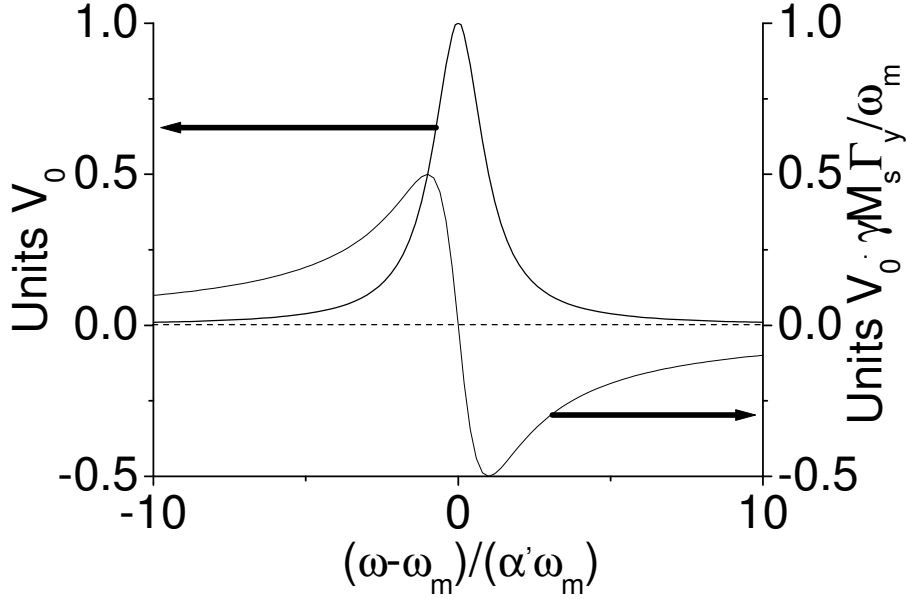}}\caption{The dimensionless DC
voltage over the sample biased by an AC current has two components that are
plotted as a function of AC current frequency; $V_{0}=I_{0}^{2}\sin^{2}%
\theta{\displaystyle \frac{\partial R(\nu)}{\partial\nu}\frac{\hbar}{2e}%
\frac{\gamma}{2M_{s}V_{m}}\frac{1}{\alpha^{\prime}\omega_{\mathrm{m}}}}$.}%
\label{FMRshape}%
\end{figure}

The resulting magnetization dynamics causes oscillations of the
magnetoresistance $R(\mathbf{m}_{1}(t),\mathbf{m}_{2}(t))=R(\cos\theta(t))$ of
the multilayer structure (details of the calculations are given in Appendix
A). In the presence of an AC current bias $I(t)=I_{0}\operatorname{Re}%
e^{i\omega t}$, the resistance has an oscillating component that in the linear
approximation reads:%
\begin{align}
R(\cos\theta(t))  &  \approx R(\nu)-\sin\theta{\frac{\partial R(\nu)}%
{\partial\nu}}\bigtriangleup\theta(t)\\
&  =R(\nu)-\sin\theta{\frac{\partial R(\nu)}{\partial\nu}}m_{y^{\prime}%
}(t)=R(\nu)-{\frac{I_{0}}{M_{s}}}\sin\theta{\frac{\partial R(\nu)}{\partial
\nu}}\operatorname{Re}(e^{i\omega t}\chi_{\mathrm{y^{\prime}I}}),
\end{align}
leading to a nonlinear phase-sensitive effect in the voltage across the
sample
\begin{align}
U  &  =R(t)I(t)=R(\nu)I(t)-\frac{\sin\theta}{4M_{s}}\frac{\partial R(\nu
)}{\partial\nu}(I_{0}e^{i\omega t}\chi_{\mathrm{y^{\prime}I}}+I_{0}e^{-i\omega
t}\chi_{\mathrm{y^{\prime}I}}^{\ast})(I_{0}e^{i\omega t}+I_{0}e^{-i\omega
t})\\
&  \sim I_{0}^{2}\operatorname{Re}[\left(  1+e^{i2\omega t}\right)
\chi_{\mathrm{y^{\prime}I}}],
\end{align}
where the parameter $\nu=\cos\theta$ describes the equilibrium configuration
of the magnetizations and $\theta(t)=\theta+\bigtriangleup\theta(t)$ (to
lowest order $\bigtriangleup\theta\approx m_{y^{\prime}}$). The magnetization
dynamics\ is thus manifest in the nonlinear response, \textit{i.e}. the zero
and second harmonic components of the voltage across the sample:%
\begin{align}
U_{0}=\frac{I_{0}^{2}\sin\theta}{2M_{s}}\frac{\partial R(\nu)}{\partial\nu
}\operatorname{Re}\chi_{\mathrm{y^{\prime}I}}(\omega)  &  \overset
{\omega\rightarrow\omega_{\mathrm{m}}}{=}-I_{0}^{2}\sin^{2}\theta
\frac{\partial R(\nu)}{\partial\nu}\frac{\hbar}{2e}\frac{\gamma}{2M_{s}V_{m}%
}\frac{1}{\omega_{\mathrm{m}}}\nonumber\\
&  \times\left(  \frac{\alpha^{\prime}\omega_{\mathrm{m}}^{2}}{(\omega
-\omega_{\mathrm{m}})^{2}+\alpha^{\prime2}\omega_{\mathrm{m}}^{2}}%
-\frac{(\omega-\omega_{\mathrm{m}})\gamma M_{s}\Gamma_{y^{\prime}}}%
{(\omega-\omega_{\mathrm{m}})^{2}+\alpha^{\prime2}\omega_{\mathrm{m}}^{2}%
}\right)  , \label{u0}%
\end{align}
\begin{align}
U_{2\omega}  &  =\frac{I_{0}^{2}\sin\theta}{2M_{s}}\frac{\partial R(\nu
)}{\partial\nu}|\chi_{\mathrm{y^{\prime}I}}(\omega)|\overset{\omega
\rightarrow\omega_{\mathrm{m}}}{=}I_{0}^{2}\sin^{2}\theta\frac{\partial
R(\nu)}{\partial\nu}\frac{\hbar}{2e}\frac{\gamma}{2V_{m}}\nonumber\\
&  \times\sqrt{\frac{1+\gamma^{2}M_{s}^{2}\Gamma_{y^{\prime}}/\omega^{2}%
}{(\omega-\omega_{\mathrm{m}})^{2}+\alpha^{\prime2}}}. \label{u2}%
\end{align}
As pointed out in Ref. \onlinecite{Tulapurkar:nat05}, the DC voltage $U_{0}$
can be interpreted as a diode-like rectification. The amplitudes $U_{0}$ and
$U_{2\omega}$ show a resonant enhancement close to $\omega_{\mathrm{m}}$ (note
that $U_{0}$ corresponds to $V_{mix}$ in Ref. \onlinecite{Sankey:prl06} and
its sign corresponds to a current flow from soft to hard ferromagnetic layer).
Eq. (\ref{u0}) is a linear combination of the symmetric and antisymmetric
Lorentzians (see Fig. \ref{FMRshape}). By fitting the DC voltage $U_{0}%
(\omega)$ to a linear combination of\ the two curves in Fig. \ref{FMRshape},
we can determine the parameter $\Gamma_{y^{\prime}}\approx\eta_{1}%
N_{x^{\prime}y^{\prime}}+\eta_{2}N_{x^{\prime}}$ that is affected by the
effective field $\eta_{2}$ and the non-collinear anisotropy $N_{x^{\prime
}y^{\prime}}$. In Fig. \ref{shape2}, we demonstrate how the resonance becomes
skewed merely by the non-collinear shape anisotropy (the spin-transfer torque 
does not lie in a plane formed by principal axes of the shape anisotropy). The resonant layer with the magnetization
$\mathbf{M}_{2}$ corresponds to the harder layer in this plot (as it was discussed in Sec. IIIA, the harder and softer layers can switch their roles with a proper choice of the AC current frequency).
The opposite scenario,\cite{Tulapurkar:nat05} without anisotropies, the effective field
$\eta_{2}$ can still cause an antisymmetric Lorentzian signal since
$\Gamma_{y^{\prime}}\approx\eta_{2}N_{x^{\prime}}=\eta_{2}H_{0}/M_{s}$, as
reported by Ref. \onlinecite{Kupferschmidt:2006}.

\subsection{Magnetovibrational resonance driven and detected by spin-transfer
torques}

\begin{figure}[ptb]
\centerline{\includegraphics[scale=0.7]{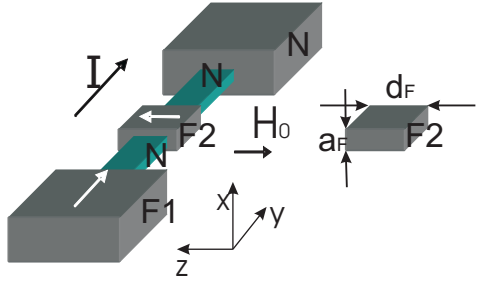}}\caption{A device to detect
magnetovibrational coupling caused by magnetic form and crystal anisotropies.
An external magnetic field $\mathbf{H}_{0}$ is used to tune the FMR frequency.
}%
\label{rev2}%
\end{figure}

In the following, we repeat the derivation of the previous section but allow
for a torsional degree of freedom along the current flow. This requires a
free-standing conducting structure, such as a doubly clamped cantilever. We
already discussed the coupled dynamics of the magnetization and the lattice
for a small magnet at the free tip of a cantilever that is clamped on the
other side.\cite{Kovalev:apl03} However, here we consider the limit in which
the magnet is heavy compared to the cantilever. This reduces the mechanical
resonance frequencies, but increases the magnetomechanical coupling strength.
The normal metal lattice serves as a spring and the ferromagnet is the load.
We first consider a well-aligned structure for which the primed and unprimed
coordinate systems in Fig. \ref{rev2} coincide. A mechanical torsion profile
described by the angle $\varphi(y)$ increases the free energy as:%
\begin{align}
F(\mathbf{M})  &  =F(\mathbf{M}_{0})+V_{m}\left(  N_{x}\left[  M_{\mathrm{x}%
}+M_{\mathrm{z}}\varphi(0)\right]  ^{2}/2+N_{y}M_{y}/2\right. \nonumber\\
&  \left.  +N_{xy}\left[  M_{\mathrm{x}}+M_{\mathrm{z}}\varphi(0)\right]
M_{y}\right)  +\frac{C}{2}\int_{-L/2}^{L/2}\left(  \frac{\partial\varphi
}{\partial y}\right)  ^{2}dy, \label{FREEapp1}%
\end{align}
where $C$ is an elastic constant defined by the shape of the cross-section and
the material of the normal metal link ($C=\mu da^{3}/3$ for a long plate with
thickness $a$ much smaller than width $d$, $a\ll d$, $\mu$ is the Lamé
constant for the normal metal), $\varphi(0)$ is the torsion angle at the
middle magnetic section. The coefficients $N_{x}$, $N_{y}$ and $N_{xy}$ have
the same meaning as in the previous section (one expects $N_{xy}=0$ in setup
of Fig. \ref{rev2}; however, we keep $N_{xy}$ in order to use our results for
more general configurations). The width $L^{\prime}$ of the magnetic layer
along the axis $y$ is supposed to be small compared to the length of the
normal metal links $L$($\gg L^{\prime}$), so internal strains and deformations
in the magnetic section are disregarded. The integration is therefore carried
out from one clamping point $y=-L/2$ to the other at $y=L/2$, excluding the
ferromagnetic layer. With Eq. (\ref{FREEapp1}), the Landau-Lifshitz-Gilbert
equation has to be modified as:
\begin{equation}
\frac{d\mathbf{M}}{dt}=-\gamma\mathbf{M}\times\mathbf{H}_{\mathrm{eff}}%
+\frac{\alpha}{M_{\mathrm{s}}}\mathbf{M}\times\left(  \frac{d\mathbf{M}}%
{dt}\right)  _{\mathrm{m}}+\gamma\frac{\hbar}{2e}\frac{I(t)}{V_{m}}\left[
\eta_{1}\mathbf{m}\times(\mathbf{m}_{\mathrm{fixed}}\times\mathbf{m})+\eta
_{2}(\mathbf{m}_{\mathrm{fixed}}\times\mathbf{m})\right]  \label{LLGapp1}%
\end{equation}%
\begin{align}
\mathbf{H}_{\mathrm{eff}}  &  =-\left(  N_{x}M_{x}+N_{xy}M_{y}+N_{x}%
M_{\mathrm{z}}\varphi(0)\right)  \mathbf{x}\\
&  -\left(  N_{y}M_{y}+N_{xy}M_{x}+N_{xy}M_{\mathrm{z}}\varphi(0)\right)
\mathbf{y},
\end{align}
where the derivative $\left(  d\mathbf{M}/dt\right)  _{\mathrm{m}}$ is defined
in the reference frame of the magnet, since most proposed mechanisms for
Gilbert damping act on the magnetization motion relative to the underlying
lattice only. This can be taken into account by
\begin{equation}
\left(  \frac{d\mathbf{M}}{dt}\right)  _{\mathrm{m}}=\frac{d\mathbf{M}}%
{dt}-\frac{d\varphi(0)}{dt}M_{\mathrm{z}}\mathbf{x}.
\end{equation}

The equation of mechanical torsional motion of the normal metal strip
is:\cite{Landau:59}%
\begin{equation}
C\frac{\partial^{2}\varphi}{\partial y{}^{2}}=\rho I\frac{\partial^{2}\varphi
}{\partial t^{2}}, \label{Mdynamics}%
\end{equation}
where $I=\int(z{}^{2}+x{}^{2})dzdx\simeq ad^{3}/12$ is again the moment of
inertia of the cross-section about its center of mass and $\rho$ is the mass
density. The oscillating solution has the form $\varphi=(A_{1}\sin
(ky)+A_{2}\cos(ky))e^{i\omega t}$, where $k=\omega/c$ is the wave number,
$c=2c_{\mathrm{t}}a/d=\sqrt{C/(\rho I)}$, and $c_{\mathrm{t}}=\sqrt{\mu/\rho}$
is the sound velocity of the transverse mode. The free constants $A_{1}$ and
$A_{2}$ are determined by the boundary conditions.

The first condition $\varphi|_{y=L/2(-L/2)}=0$ corresponds to perfect clamping
at the boundaries. The second one, at the connection with the ferromagnetic
load, can be obtained from the variational principle applied to Eq.
(\ref{FREEapp1}), and corresponds to the torque $C\partial\varphi/\partial
y|_{y=\pm L^{\prime}/2}$ exerted by the magnetization at the interface to the
normal metal link:%
\begin{equation}
C\frac{\partial\varphi}{\partial y}|_{y=\pm L^{\prime}/2}=\frac{1}{\gamma
}\left(  \frac{d\mathbf{M}}{dt}+\gamma\mathbf{M}\times\mathbf{H}_{0}\right)
|_{y}+L^{\prime}I_{F}\rho_{F}\frac{d^{2}\varphi(0)}{dt^{2}}, \label{bound}%
\end{equation}
where $I_{F}$ is the moment of inertia of the magnetic load (for a thin plate
of mass $M$, we can approximate it as $L^{\prime}I_{F}\rho_{F}=Md_{F}^{2}/12,$
where $d_{F}$ is its width, see Fig. \ref{rev2}), and $\partial\varphi
/\partial y|_{y=\pm L^{\prime}/2}$ is the derivative of the torsion angle of
the normal metal link at the connection with the ferromagnet. This boundary
condition is equivalent to the conservation of the mechanical angular momentum
written for the magnetic load. With $C\partial\varphi/\partial y|_{y=\pm
L^{\prime}/2}=Ck\varphi\cot kL$, we obtain:%
\begin{equation}
Ck\varphi(0)\cot(kL)-L^{\prime}I_{F}\rho_{F}\frac{d^{2}\varphi(0)}{dt^{2}%
}=\frac{1}{\gamma}\left(  \frac{d\mathbf{M}}{dt}+\gamma\mathbf{M}%
\times\mathbf{H}_{0}\right)  |_{y}, \label{Fourier}%
\end{equation}
where the right hand side describes the magnetically induced torques that to
leading order equal the Gilbert and anisotropy torques. In the absence of the
magnetic anisotropies and Gilbert damping, the two terms ${d\mathbf{M/}dt}$
and $\gamma\mathbf{M}\times\mathbf{H}_{0}$ cancel each other.

By only considering the left hand side of Eq. (\ref{Fourier}), we can find the
mechanical resonance frequency as $\omega_{\mathrm{e}}=\sqrt{Ck\cot
(kL)/(L^{\prime}I_{F}\rho_{F})}\approx\sqrt{C/(LL^{\prime}I_{F}\rho_{F})}$,
where the approximation holds when $\omega_{\mathrm{e}}$ is smaller than the
resonant frequency of the normal metal link with a free end (which corresponds
to the heavy-load limit). In frequency space, the left hand side of Eq.
(\ref{Fourier}) can be expressed in terms of the dimensionless response
function $F(\omega)$ of the mechanical subsystem to an oscillating torque of
frequency $\omega$ applied to the load:%
\begin{equation}
F(\omega)=\omega_{\mathrm{e}}^{2}/\left[  (\omega^{2}-\omega_{\mathrm{e}}%
^{2}+2i\beta\omega)\right]
\end{equation}
where $\beta$ is a phenomenological damping constant describing dissipation in
Eq. (\ref{Mdynamics}) and it is related to the quality factor $Q$ of
oscillator at the resonance frequency $\omega_{\mathrm{e}}$ as $Q=\omega
_{\mathrm{e}}/(2\beta)$ (at 1 GHz $Q\sim500$).\cite{Roukes:nat03} In terms of
$F(\omega)$, the mechanical response function reads:%
\begin{equation}
(\varphi/T)_{\omega}=\left(  1/Ck\cot(kL)\right)  F(\omega)\approx
(L/C)F(\omega), \label{Mrespons}%
\end{equation}
where $T$ is the torque externally exerted on the load.

To first order in the mechanical and magnetization oscillations, Eqs.
(\ref{LLGapp1}) and (\ref{Fourier}) lead us to the following response
function:%
\begin{align}
{\chi_{\mathrm{yI}}(\omega)}  &  {=}(M_{y}/I)_{\omega}={\frac{\hbar\gamma
}{2eV_{m}}}\nonumber\\
&  {\times\frac{-i\eta_{1}\omega+\Gamma_{y}\gamma\left[  M_{s}+g\left(
H_{0}/N_{x}\right)  F(\omega)\right]  }{\omega^{2}-\omega_{\mathrm{m}}%
^{2}+2i\alpha^{\prime}\omega\omega_{\mathrm{m}}+\Lambda F(\omega)}%
}\label{MVR1}\\
&  \overset{g\rightarrow0}{\approx}{\frac{\hbar\sin\theta\gamma}{2eV_{m}}%
\frac{-i\eta_{1}\omega+\gamma M_{s}\Gamma_{y}}{\omega^{2}-\omega_{\mathrm{m}%
}+2i\alpha^{\prime}\omega\omega_{\mathrm{m}}+\Lambda F(\omega)},}%
\end{align}
where%
\begin{equation}
\Lambda=g\left[  \omega^{2}-H_{0}\omega_{\mathrm{m}}/(N_{x}M_{s})\right]  ,
\label{ReG}%
\end{equation}
and
\begin{align}
g  &  =M_{\mathrm{s}}^{2}V_{m}N_{x}\left(  1/Ck\cot(kL)\right) \\
&  \approx M_{\mathrm{s}}^{2}V_{m}N_{x}L/C=N_{x}(L/a)^{2}(V_{m}/V_{l}%
)(M_{\mathrm{s}}^{2}/\mu),\nonumber
\end{align}
$V_{m}$ and $V_{l}$ are the volumes of the magnetic load and the normal metal
spring, respectively. $L$ and $a$ are the largest and the smallest dimensions
of the normal metal links, $k=\omega_{\mathrm{e}}/c$. We recover here the
results from Ref. \onlinecite{Kovalev:jjap06}, implying that the spin
polarized current is equivalent to an external rf field along the $x$-axis
applied to a magnetic cantilever. Since the ratio $V_{m}/V_{l}$ can be made
much larger compared to the limit of light load considered in Ref.
\onlinecite{Kovalev:jjap06}, we conclude that by making the normal metal links
thinner (in setup of Ref. \onlinecite{Kovalev:jjap06} we have to make the
cantilever thinner) we reduce the stiffness of the device, which results in a
better sensitivity and stronger coupling. The reduced stiffness leads to a
drop in the resonance frequency that in turn can be compensated by making the
structures smaller.

\begin{figure}[ptb]
\centerline{\includegraphics{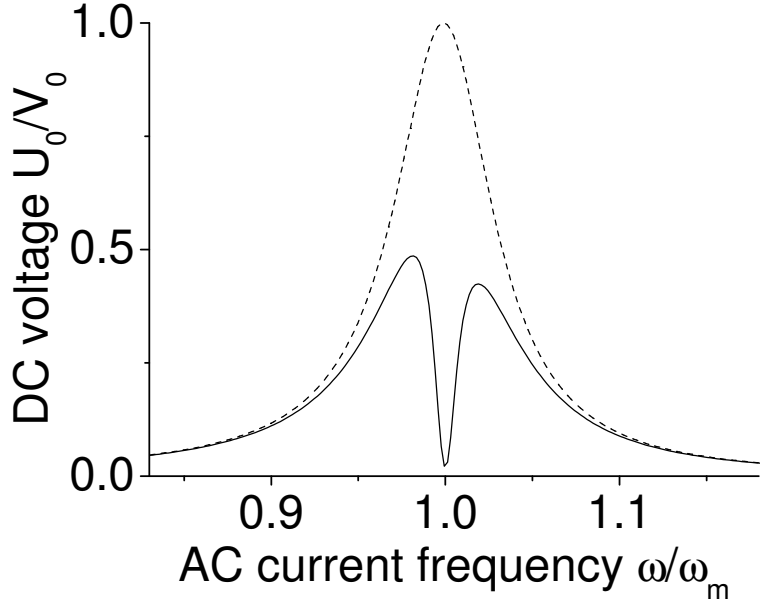} }\caption{Dependence of the DC
component of the voltage on the frequency or the AC current bias for purely
FMR (dashed line) and MVR (solid line) ($\omega_{m}=\omega_{e}$,
$\alpha^{\prime}=0.02$, $\beta/\omega=0.002$, $g=0.001$).}%
\label{MMR0}%
\end{figure}

\begin{figure}[ptb]
\centerline{\includegraphics{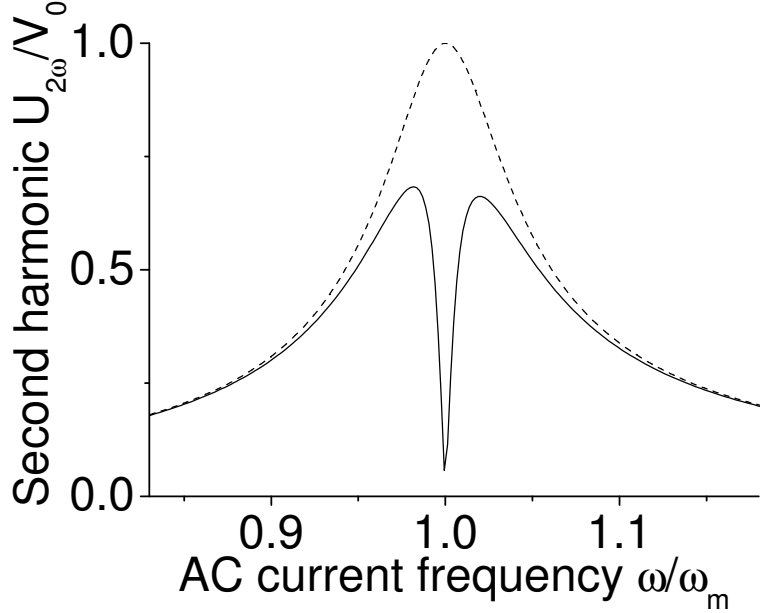} }\caption{Dependence of the second
harmonic of the voltage on the frequency of the AC current bias for\ purely
FMR (dashed line) and MVR (solid line) ($\omega_{m}=\omega_{e}$,
$\alpha^{\prime}=0.02$, $\beta/\omega=0.002$, $g=0.001$).}%
\label{MMR2}%
\end{figure}

The \textit{nonlinear} response to an AC current can now be found by
substituting Eq. (\ref{MVR1}) into the first parts of Eqs. (\ref{u0}%
,\ref{u2}). We conclude that the most conspicuous feature of the
magnetovibrational coupling\ is the formation of a \textit{magnetopolariton}
and the splitting of the ferromagnetic resonance close to $\omega
=\omega_{\mathrm{m}}$, which is governed by $\sqrt{\Lambda}$. The expression
for $\Lambda,$ Eq. (\ref{ReG}), suggests that the splitting can be tuned by
the external magnetic field. The line width of those two resonances is defined
by $\alpha^{\prime}$ and $\beta$, and the shape is a combination of the
symmetric and antisymmetric Lorentzians as in Fig. \ref{FMRshape}. Let us make
estimates for a system shown in Fig. \ref{rev2}. The dimensions of the
metallic links are here chosen as $\left(  0.5\times0.05\times0.01\right)
\operatorname{\mu m}%
$ with a Py load of the size $\left(  0.1\times0.2\times0.02\right)
\operatorname{\mu m}%
$, for which we can estimate a resonance frequency in the range of
$0.5\text{GHz}$ ($\mu\sim100\mathrm{G}%
\operatorname{Pa}%
$) \cite{Simmons:75} and a coupling parameter $g\sim0.001$ (free-standing
metallic structures of such dimensions have already been
realized\cite{Paraoanu:apl05}). For intermetallic interfaces, the{}%
\textquotedblleft effective field effect\textquotedblright\ due to $\eta_{2}$
is very small, thus the phase shift of the magnetization is absent
($\Gamma_{y}=0$). The amplitudes, given by Eqs. (\ref{u0},\ref{u2}), are large
in the proximity of the ferromagnetic resonance (FMR). The magnetovibrational
coupling splits the FMR peak by $\sqrt{g}\omega$ ($H_{0}<N_{x}M_{s}$), which
allows an electric detection of the mechanical motion excited by the spin
transfer. Normalized voltages are plotted in Figs. \ref{MMR0} and \ref{MMR2}
for purely ferromagnetic (mechanical motion is suppressed) compared to
magneto-vibrational (MVR) resonances.

Finally, we write the response function for arbitrary magnetization directions
in Fig. (\ref{rev1}):%
\begin{align}
{\chi_{\mathrm{y^{\prime}I}}(\omega)}  &  {=}(M_{y^{\prime}}/I)_{\omega
}={\frac{\hbar\gamma\sin\theta}{2eV_{m}}}\label{MVR11}\\
&  \overset{g\rightarrow0}{\approx}{\frac{\hbar\sin\theta\gamma}{2eV_{m}}%
\frac{-i\eta_{1}\omega+\gamma M_{s}\Gamma_{y^{\prime}}}{\omega^{2}%
-\omega_{\mathrm{m}}+2i\alpha^{\prime}\omega\omega_{\mathrm{m}}+\Lambda
F(\omega)},}%
\end{align}
where $\Lambda$ is defined in Eq. (\ref{ReG1}). As one can see, the form of
the response function does not change. Eq. (\ref{MVR11}) can be substituted
into the first parts of Eqs. (\ref{u0},\ref{u2}) in order to calculate the
\textit{nonlinear} response to an AC current. Similarly to Eq. (\ref{damping1}%
), the parameter $\Gamma_{y^{\prime}}\approx\eta_{1}N_{x^{\prime}y^{\prime}%
}+\eta_{2}N_{x^{\prime}}$ governs the balance between the symmetric and
antisymmetric Lorentzians composing each resonance (see Fig. \ref{FMRshape}).
The magnetovibrational coupling is described by $\Lambda$:
\begin{align}
\operatorname{Re}\Lambda &  =g\left[  \omega^{2}-(H_{z^{\prime}}+\cot
\theta_{1}H_{y^{\prime\prime}})\omega_{\mathrm{m}}/(N_{x^{\prime\prime}}%
M_{s})\right] \label{ReG1}\\
\operatorname{Im}\Lambda &  =g\omega\gamma\left[  H_{z^{\prime}}%
N_{x^{\prime\prime}y^{\prime\prime}}/N_{x^{\prime\prime}}+\cot\theta
_{1}(H_{x^{\prime\prime}}+H_{y^{\prime\prime}}N_{x^{\prime\prime}%
y^{\prime\prime}}/N_{x^{\prime\prime}})\right]
\end{align}
and
\begin{align}
g  &  =\sin^{2}\theta_{1}M_{\mathrm{s}}^{2}V_{m}N_{x^{\prime\prime}}\left(
1/Ck\cot(kL)\right) \\
&  \approx\sin^{2}\theta_{1}M_{\mathrm{s}}^{2}V_{m}N_{x^{\prime\prime}%
}L/C=\sin^{2}\theta_{1}N_{x^{\prime\prime}}(L/a)^{2}(V_{m}/V_{l}%
)(M_{\mathrm{s}}^{2}/\mu),\nonumber
\end{align}
where $V_{m}$ and $V_{l}$ are the volumes of the magnetic load and the normal
metal spring, respectively. $L$ and $a$ are the largest and the smallest
dimensions of the normal metal links, $k=\omega_{\mathrm{e}}/c$, $\theta_{1}$
is the angle between the equilibrium direction of the magnetization
$\mathbf{M}_{2}$ and the current flow, and the external field is
$\mathbf{H}_{0}=(H_{x^{\prime\prime}},H_{y^{\prime\prime}},H_{z^{\prime}})$. The
parameter $\Lambda$ is calculated in a reference frame $x^{\prime\prime
},y^{\prime\prime},z^{\prime}$ shown in Fig. (\ref{rev1}). The $x^{\prime
\prime}$-axis is perpendicular to the current flow and the axis $z^{\prime}$
(in general, this reference frame is different from the one used in Sec. IIIA,
since the $x^{\prime\prime}$-axis is not necessarily along the direction of
$\mathbf{m}_{1}\times\mathbf{m}_{2}$).

\subsection{Mechanical torques due to absorption of transversely polarized
currents}

Let us now apply a DC current bias to the system in Fig. \ref{rev2}. According
to the magnetoelastic equations discussed above, the angular momentum of the
spin polarized current is completely transformed into mechanical torques when
the crystal and shape anisotropies are strong enough to prevent magnetization
motion relative to the lattice. The mechanical torque then equals the
spin-transfer torque:%
\begin{equation}
T_{st}=-\gamma\frac{\hbar}{2e}I\eta_{1}\mathbf{m}\times(\mathbf{m}%
_{\mathrm{fixed}}\times\mathbf{m})\approx-\gamma T_{0}\eta_{1}%
,\label{STtorque}%
\end{equation}
The Gilbert damping does not appear explicitly in this formula; however, it
determines the time scale $1/(\alpha\gamma M_{s})$ in which the system reaches
quasi-equilibrium. The upper boundary for the spin-transfer-induced mechanical
torque for a thin film of the size $\left(  20\times200\times200\right)
\operatorname{nm}$ (without crystal anisotropy) is defined by the maximum
effective magnetic field due to the form anisotropy, $H_{eff}=4\pi\theta
M_{\mathrm{s}}$, where $\theta\approx\pi/20\sim H_{0}/M_{\mathrm{s}}$
corresponds to some small stationary deflection of the magnetization out of
the plane of the film in the presence of the stabilizing external magnetic
field $\mathbf{H}_{0}$ (Fig. \ref{rev2}). This field generates a torque of
$\gamma4\pi\theta M_{\mathrm{s}}\cdot M_{\mathrm{s}}V_{m}/\gamma\sim
10^{-16}\operatorname{N}\operatorname{m}$, where $V_{m}$ is the volume of the
ferromagnet and $M_{\mathrm{s}}=10^{6}\
\operatorname{A}%
/%
\operatorname{m}%
$ is the saturation magnetization for Py. Such torques are well above the
sensitivity of existing NEMS oscillators.\cite{Huang:njp05}

\section{Spin-transfer nanomotor}

\begin{figure}[ptb]
\centerline{\includegraphics[scale=0.5]{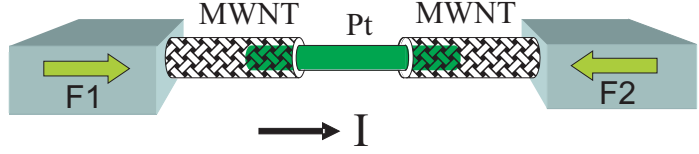}}\caption{A spin-transfer
nanomotor in which a metallic wire (rotor) with strong spin-flip scattering
(such as Pt) is grown inside a MWNT that is connected to two ferromagnetic
contacts.}%
\label{rev3}%
\end{figure}

Finally, we address the question how the torque can be transformed into a
potentially useful rotary motion. It has been suggested in the literature to
use carbon nanotubes as bearings for metallic
nanowires.\cite{Kral:prb02,Fennimore:nat03} In Fig. \ref{rev3}, we propose a
design of a spin-transfer nanomotor based on multi-wall carbon nanotube (MWNT)
connected to two ferromagnetic electrodes (the torque doubles when the second
electrode is ferromagnetic, but one ferromagnet is sufficient in principle). A
metallic nanowire with strong spin-flip scattering nanowire is encapsulated by
the MWNT. Pt would be a good choice ($l_{sd}^{Pt}\sim20\operatorname{nm}$ at
4.2 K\cite{Olson:apl05} and $l_{sd}^{Pt}\sim1\operatorname{nm}$ at room
temperature\cite{Mizukami:prb02}). FeCo nanowires that have already been grown
inside nanotubes,\cite{Elias:nl05} presumably have spin-flip diffusion lengths
not much different from FeNi ($l_{sd}^{Py}\sim5\operatorname{nm}%
)$,\cite{Urazhdin} which makes this material also very suitable. The metallic
nanowire should preferably be longer than the spin-diffusion length in order
to achieve a complete angular momentum transfer. After burning off the outer
shells over the platinum nanowire (in addition, the MWNT may also be pulled
out to open the Pt wire), we force the current to flow through the metallic
wire that serves as a spin-sink and a rotor. It has recently been calculated
that the conductance may rapidly oscillate due to quantum interference effects
as we change the overlap between two nanotube
shells,\cite{Kim:prb02,Tamura:prb05,Tunney:cm06} but disorder strongly enhance
the intershell conductance, consistent with experiments.\cite{Cumings:prl04}

The situation with not too high polarization of the MWNT connection to
ferromagnet\cite{Tsukagoshi:nat99} ($P\sim0.01$) has tendency to improve with
recent experiments reporting $\mbox{TMR}=5\%$ ($P\sim0.2$%
).\cite{Sahoo:nat05,Tombros:prb06,Man:prb06} Half-metallic contacts to
nanotubes might lead to much higher values.\cite{Hueso:apl06} Adopting the
maximum current through a single MWNT measured to date\cite{Dohn:sl02}
$I=1\operatorname{mA}$, we arrive at an estimate for the mechanical torque
generated in the rotor $T\sim T_{0}\simeq10^{-19}\operatorname{N}%
\operatorname{m}$ where we optimistically assume $P\sim1$. This exceeds by
many orders of magnitude the torques that can be induced by circularly
polarized light ($T\sim10^{-29}\operatorname{N}\operatorname{m}$%
).\cite{Kral:prb02} The advantage of a MWNT bearing is that the friction force
can be very small. The bearing may still get stuck at certain preferred
positions that minimize the interaction energy of the sleeve and shaft.
However, each layer is likely to have a different chirality, so the potential
barrier (static friction force) hindering rotation should be very small. An
upper boundary for the static friction force per unit area has been
measured\cite{Cumings:sc00} to be $6.6\times10^{-13}%
\operatorname{N}%
/\operatorname{nm}^{2}$ (but the actual value is possibly much smaller). For a
nanotube radius of $2\operatorname{nm}$ this corresponds to a static friction
torque per unit area of less than $10^{-21}\operatorname{N}\operatorname{m}%
/\operatorname{nm}^{2}$. For the same nanotube radius, the overlap length
between the rotor and the outer shells can be made at least
$10\operatorname{nm}$ without the risk that the rotor gets stuck by the static
friction. When the barrier to rotation is overcome, the static friction is
taken over by the dynamic friction. The latter is much smaller initially, but
increases proportional to the angular velocity.\cite{Servantie:cm06}

The proposed motor could be useful in the next generation of synthetic
nanometer-scale electromechanical systems. With an attached metal plate the
rotor can serve as a mirror, with relevance to high-density switching
devices.\cite{Fennimore:nat03} The motor can find applications for inducing
and detecting motion in microfluidics systems and biological systems.

\section{Conclusions and outlook}

We find that the electric current-induced creation and detection of mechanical
torques in magnetic nanostructures is possible by rectification technique
suggested recently in Ref. \onlinecite{Tulapurkar:nat05} and
\onlinecite{Sankey:prl06}. We develop a detailed theory of the magnetization
dynamics in the presence of electric currents relevant for these experiments
that agrees with the very recent study by Kupferschmidt \textit{et
al}..\cite{Kupferschmidt:2006} \ Subsequently, we predict that an alternating
current can drive magnetovibrational dynamics, which can be read out by the
generated DC voltage. Finally, we come to the conclusion that electric
current-induced mechanical torques can create rotary motion. We propose a
novel spin-transfer driven nanomotor based on integrating metallic nanowires
with carbon nanotubes.

In the study of the rectification effect, we find that the DC voltage
originates from two mechanisms: the rectification of the applied AC current
and the spin-pumping by the precessing ferromagnet. The second mechanism is
important only in very thin ferromagnetic layers when the Gilbert damping is
strongly enhanced (in accord with Ref. \onlinecite{Kupferschmidt:2006}). It
should be noted that the shape of DC voltage as a function of frequency is a
symmetric Lorentzian for the spin-pumping mechanism. This can be distinguished
from the voltage induced in the presence of noncollinear magnetic anisotropies
or the {}{}``effective\textquotedblright\ spin-transfer field, which causes
asymmetric line shapes.

We generalized these results to treat magnetovibrationally coupled systems.
The strongest coupling is achieved when the lowest mechanical mode is at
resonance with the FMR frequency. In that case, a magnetopolariton is formed,
and the Lorentzian shape of the DC voltage splits\ by an amount that is
governed by the sensitivity of the mechanical system to external torques and
by the magnetic anisotropies (without which there would be no
magnetovibrational coupling). The technique based on the resonant
magnetovibrational coupling can also be used to detect vibrations that are
created externally. It can therefore be an alternative to the magnetomotive
technique employed in fast transducers of mechanical motion.\cite{Huang:njp05}

We conclude that the functionalities of the spin-transfer torque, already used
in applications such as magnetic memories, can be extended by taking into
account the coupling with the mechanical degrees of freedom. The experimental
realization is a challenge, since free-standing metallic small structures on
micro and nanoscale need to be fabricated and manipulated.

\begin{acknowledgments}
We thank Yaroslav Tserkovnyak, Xuhui Wang, and Yuli Nazarov for stimulating
discussions. We appreciate helpful communications with Dan Ralph and Jack
Sankey. We are grateful that Piet Brouwer sent us manuscript (Ref.
\onlinecite{Kupferschmidt:2006}), reminding us of the importance of spin
pumping. This work has been supported by the Dutch FOM Foundation, National
Science Foundation Grant No. PHY99-07949, the EU Commission FP6 NMP-3 project
505587-1{}{}{}``SFINX\textquotedblright, and the Research Council of Norway
Grant No. 162742/V00.
\end{acknowledgments}

\appendix

\section{Effect of mixing conductance on the torque and magnetoresistance}

In this Appendix, we generalize the relations derived in Ref.
\onlinecite{Kovalev:prb02} for the torques and resistances in a general N%
$\vert$%
F1%
$\vert$%
N%
$\vert$%
F2%
$\vert$%
N multilayered structures without bulk layer spin-flip scattering to the
presence of an imaginary part of the mixing conductances or{}{}``effective
field\textquotedblright. We use magnetoelectronic circuit theory as formulated
by Eqs. (\ref{chargebody},\ref{spinbody}). Our result for the spin-transfer
torque acting on the first ferromagnet (the torque on the second ferromagnet
can be obtained by permutation of indexes $1$ and $2$, and by replacing $I$ with $-I$) can be summarized by:%
\begin{equation}
\mathbf{T}_{ST}^{1}=\frac{I\hbar}{2e}\left[  \eta_{s}\mathbf{m}_{1}%
\times(\mathbf{m}_{1}\times\mathbf{m}_{2})+\eta_{f}(\mathbf{m}_{2}%
\times\mathbf{m}_{1})\right]  , \label{torqMix}%
\end{equation}
where we introduced the spin-transfer efficiencies:%
\[%
\begin{array}
[c]{ll}%
\eta_{s}= & \left(  \left[  R_{2-}(R_{r}+R_{1})-R_{1-}R_{2}\alpha\right]
(1+{\displaystyle \frac{R_{2r}}{R_{r}}}\widetilde{G}_{1i}^{2})\right. \\
& \left.  \left.  +{\displaystyle \frac{R_{1r}}{R_{r}}\left[  \frac{R_{2-}%
}{G_{1}}-\frac{R_{1-}}{G_{2}}\alpha\right]  \widetilde{G}_{2i}^{2}%
}+{\displaystyle \frac{R_{1r}R_{2r}}{R_{r}}}(R_{1-}+R_{2-}\alpha)\widetilde
{G}_{1i}\widetilde{G}_{2i}\right)  \right/ \\
& \left(  (R_{r}+R_{1})(R_{r}+R_{2})-R_{1}R_{2}\alpha^{2}%
+{\displaystyle \left[  \frac{(R_{r}+R_{1})}{G_{2}}-\frac{R_{2}}{G_{1}}%
\alpha^{2}\right]  \frac{R_{2r}}{R_{r}}}\widetilde{G}_{1i}^{2}\right. \\
& \left.  {\displaystyle +\left[  \frac{(R_{r}+R_{2})}{G_{1}}-\frac{R_{1}%
}{G_{2}}\alpha^{2}\right]  \frac{R_{1r}}{R_{r}}\widetilde{G}_{2i}^{2}}%
+2\alpha{\displaystyle \frac{R_{1r}R_{2r}}{R_{r}}}({\displaystyle \frac
{1}{G_{1}}}+{\displaystyle \frac{1}{G_{2}}})\widetilde{G}_{1i}\widetilde
{G}_{2i}\right)
\end{array}
,
\]
\[%
\begin{array}
[c]{ll}%
\eta_{f}= & \left(  \left[  R_{2-}\left(  R_{r}+R_{1}\right)  -R_{1-}%
R_{2}\alpha\right]  {\displaystyle \frac{R_{1r}}{R_{r}}}\widetilde{G}%
_{1i}-\left[  R_{1-}\left(  R_{r}+R_{2}\right)  -R_{2-}R_{1}\alpha\right]
{\displaystyle \frac{R_{2r}}{R_{r}}}\widetilde{G}_{2i}\right. \\
& \left.  \left.  +{\displaystyle \frac{R_{1r}}{R_{r}}}{\displaystyle \left[
\frac{R_{2-}}{G_{1}}-\frac{R_{1-}}{G_{2}}\alpha\right]  \widetilde{G}%
_{1i}\widetilde{G}_{2i}^{2}}-{\displaystyle \frac{R_{2r}}{R_{r}}%
}{\displaystyle \left[  \frac{R_{1-}}{G_{2}}-\frac{R_{2-}}{G_{1}}%
\alpha\right]  \widetilde{G}_{2i}\widetilde{G}_{1i}^{2}}\right)  \right/ \\
& \left(  (R_{r}+R_{1})(R_{r}+R_{2})-R_{1}R_{2}\alpha^{2}%
+{\displaystyle \left[  \frac{(R_{r}+R_{1})}{G_{2}}-\frac{R_{2}}{G_{1}}%
\alpha^{2}\right]  \frac{R_{2r}}{R_{r}}}\widetilde{G}_{1i}^{2}\right. \\
& \left.  {\displaystyle +\left[  \frac{(R_{r}+R_{2})}{G_{1}}-\frac{R_{1}%
}{G_{2}}\alpha^{2}\right]  \frac{R_{1r}}{R_{r}}\widetilde{G}_{2i}^{2}}%
+2\alpha{\displaystyle \frac{R_{1r}R_{2r}}{R_{r}}}({\displaystyle \frac
{1}{G_{1}}}+{\displaystyle \frac{1}{G_{2}}})\widetilde{G}_{1i}\widetilde
{G}_{2i}\right)
\end{array}
,
\]
with $\alpha=\cos\theta,$ $4R_{1(2)}=1/G_{1(2)\uparrow}+1/G_{1(2)\downarrow
}-2/G_{1(2)r}$, $4R_{1(2)-}=1/G_{1(2)\uparrow}-1/G_{1(2)\downarrow}$,
$4/G_{1(2)}=1/G_{1(2)\uparrow}+1/G_{1(2)\downarrow}$, $2R_{1(2)r}=1/G_{1(2)r}%
$, $2R_{r}=1/G_{1r}+1/G_{2r}$ and $\widetilde{G}_{1(2)i}=G_{1(2)i}/G_{1(2)r}$,
where $G_{1(2)\uparrow}$ and $G_{1(2)\downarrow}$ are conductances of the left
(right) ferromagnet including the left (right) normal layer (when necessary,
the middle layer conductance can also be included in these conductances),
$G_{1(2)\uparrow\downarrow}=G_{1(2)r}+iG_{1(2)i}$ is the mixing conductance of
the left (right) ferromagnet.

The angular magnetoresistance reads:%
\begin{equation}%
\begin{array}
[c]{ll}%
R(\theta)= & R_{r}+R_{1}+R_{2}-\left(  \left[  R_{1-}^{2}+R_{2-}^{2}%
+2R_{1-}R_{2-}\alpha\right]  (1+{\displaystyle {\displaystyle \frac{R_{2r}%
^{2}}{R_{r}^{2}}}\widetilde{G}_{1i}^{2}+\frac{R_{1r}^{2}}{R_{r}^{2}}%
}\widetilde{G}_{2i}^{2})\right. \\
& +{\displaystyle \frac{(1-\alpha^{2})}{R_{r}^{2}}}\left(  {\displaystyle \left[
R_{1-}^{2}R_{2}+R_{2-}^{2}R_{1}\right]  \left[
R_{r}+{\displaystyle {\displaystyle R_{2r}}\widetilde{G}_{1i}%
^{2}+R_{1r}}\widetilde{G}_{2i}^{2}\right]  }+R_{1-}^{2}%
R_{1r} R_{2r}\widetilde{G}_{2i}^{2}+R_{2-}^{2}R_{1r} R_{2r}\widetilde{G}_{1i}^{2}\right) \\
& \left.  \left.  +2(R_{1-}+R_{2-}\alpha)(R_{2-}+R_{1-}\alpha
){\displaystyle \frac{R_{1r}R_{2r}}{R_{r}^{2}}}\widetilde{G}_{1i}\widetilde
{G}_{2i}\right)  \right/ \\
& \left(  (1+{\displaystyle \frac{R_{1}}{R_{r}}})(R_{r}+R_{2}%
)-{\displaystyle \frac{R_{1}}{R_{r}}}R_{2}\alpha^{2}+{\displaystyle \left[
\frac{(R_{r}+R_{1})}{G_{2}}-\frac{R_{2}}{G_{1}}\alpha^{2}\right]  \frac
{R_{2r}}{R_{r}^{2}}}\widetilde{G}_{1i}^{2}\right. \\
& \left.  {\displaystyle +\left[  \frac{(R_{r}+R_{2})}{G_{1}}-\frac{R_{1}%
}{G_{2}}\alpha^{2}\right]  \frac{R_{1r}}{R_{r}^{2}}\widetilde{G}_{2i}^{2}%
}+2\alpha{\displaystyle \frac{R_{1r}R_{2r}}{R_{r}^{2}}}({\displaystyle \frac
{1}{G_{1}}}+{\displaystyle \frac{1}{G_{2}}})\widetilde{G}_{1i}\widetilde
{G}_{2i}\right)
\end{array}
\label{MresMix}%
\end{equation}

\begin{figure}[ptb]
\centerline{\includegraphics{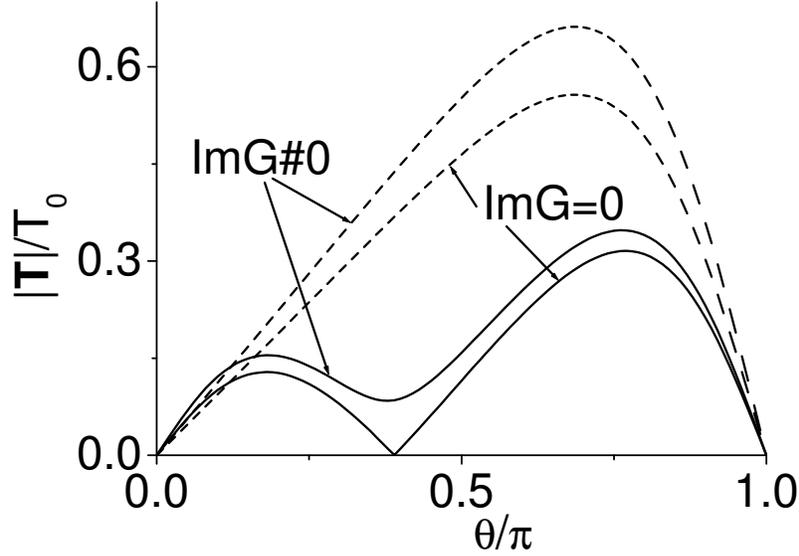} }\caption{Enhancement of the torque
by a nonzero imaginary part of the spin-mixing conductance; dashed line for
symmetric ($P=R_{-}G=0.7$, $G/G_{r}=0.6$, $\widetilde{G}_{i}=0.6$) and solid
line for asymmetric junctions ($P_{1}=R_{1-}G_{1}=0.6$, $G_{1}/G_{1r}%
=G_{1}/G_{2r}=0.6$, $\widetilde{G}_{1i}=-\widetilde{G}_{2i}=0.4$, $G_{1}%
=2G_{2}$, $P_{2}=R_{2-}G_{2}=0.13$).}%
\label{EffMix}%
\end{figure}

When the imaginary part of the mixing conductance is small, the absolute value
of the torque Eq. (\ref{torqMix}) and the magnetoresistance Eq. (\ref{MresMix}%
) contain only second order corrections in $\widetilde{G}_{1i}$ and
$\widetilde{G}_{2i}$. In general, we conclude that the torque is enhanced when
the imaginary part is not zero, as it is shown in Fig. \ref{EffMix}. This
enhancement makes it impossible to have noncollinear points of zero torque
reported earlier \cite{Kovalev:prb02,Manschot:prb04} (see Fig. \ref{EffMix})
which can influence the stable precessional states.

\section{Spin pumping and DC voltage in asymmetric N%
$\vert$%
F%
$\vert$%
N structures}

We consider here an asymmetric N1%
$\vert$%
F%
$\vert$%
N2 layered structure connected to two reservoirs and excited by rf magnetic
fields. We are interested in the DC voltage that builds up under the condition
of zero charge current. The system can be described by the generalized LLG
equation and magnetoelectronic circuit theory:%
\begin{equation}
\frac{d\mathbf{M}}{dt}=-\gamma\mathbf{M}\times\mathbf{H}_{\mathrm{eff}}%
+\frac{\alpha}{M_{\text{s}}}\mathbf{M}\times\left(  \frac{d\mathbf{M}}%
{dt}\right)  -\frac{\gamma\hbar}{2eV_{m}}\mathbf{m}\times\left[
(\mathbf{I}_{s1}+\mathbf{I}_{s2})\times\mathbf{m}\right]  \label{LLGpump}%
\end{equation}
\begin{equation}
I_{1(2)}=(G_{\uparrow}+G_{\downarrow})(\mu_{0}^{2(1)}-\mu_{0}^{1(2)}%
)+(G_{\uparrow}-G_{\downarrow})(\mathbf{\boldsymbol{\mu}}_{s}^{2(1)}%
-\mathbf{\boldsymbol{\mu}}_{s}^{1(2)})\cdot\mathbf{\mathbf{m}} \label{charge}%
\end{equation}
\begin{align}
\mathbf{I}_{s1(2)}  &  =\mathbf{\mathbf{m}}\left[  (G_{\uparrow}%
-G_{\downarrow})(\mu_{0}^{2(1)}-\mu_{0}^{1(2)})+(G_{\uparrow}+G_{\downarrow
})(\mathbf{\boldsymbol{\mu}}_{s}^{2(1)}-\mathbf{\boldsymbol{\mu}}_{s}%
^{1(2)})\right] \nonumber\\
&  -\left(  2\mathbf{\mathbf{m}}\times\mathbf{\boldsymbol{\mu}}_{s}%
^{1(2)}-\hbar\dot{\mathbf{m}}\right)  \times\mathbf{\mathbf{m}}G_{r}-\left(
2\mathbf{\mathbf{m}}\times\mathbf{\boldsymbol{\mu}}_{s}^{1(2)}-\hbar
\dot{\mathbf{m}}\right)  G_{i}, \label{spin}%
\end{align}
where $G_{\uparrow}$ and $G_{\downarrow}$ describe the spin-dependent
conductance limited by interface and bulk scattering of the ferromagnet,
$G_{\uparrow\downarrow}=G_{r}+iG_{i}$ is the interface mixing conductance of
the ferromagnet, the vector $\mathbf{\mathbf{m}}=\mathbf{M}/M_{\mathrm{s}}$ is
the direction of the magnetization, $\mu_{0}$ and $\mathbf{\boldsymbol{\mu}%
}_{s}$ are the chemical potential and spin-accumulation in the normal metals, respectively.

For metallic interfaces, $G_{i}$ usually amounts to only a few percent of
$G_{r}$. We calculate here the linear response to the external rf magnetic
field when $G_{i}$ is disregarded:%
\begin{equation}
\chi_{\mathrm{xx}}(\omega)=(M_{x}/h_{x})_{\omega}=\frac{(\gamma M_{s}%
)^{2}(N_{y}+N_{xy}\alpha)}{\omega^{2}(1+\alpha_{se}^{2})-\omega_{\mathrm{m}%
}^{2}(1+\alpha^{2})+2i\alpha^{\prime}\omega\omega_{\mathrm{m}}}.
\label{suspumpy}%
\end{equation}
\begin{equation}
\chi_{\mathrm{yx}}(\omega)=(M_{y}/h_{x})_{\omega}=\frac{\gamma M_{s}\left[
\gamma M_{s}(N_{xy}+N_{x}\alpha)-i\omega\right]  }{\omega^{2}(1+\alpha
_{se}^{2})-\omega_{\mathrm{m}}^{2}(1+\alpha^{2})+2i\alpha^{\prime}\omega
\omega_{\mathrm{m}}}. \label{suspumpx}%
\end{equation}
where $\alpha$ is the bulk Gilbert damping, $\alpha_{se}=\gamma\hbar
^{2}\left(  G_{1}+G_{2}\right)  /\left(  2eM_{s}V\right)  $ is the extra
damping due to spin-emission, $1/G_{1}=1/g_{1}+1/2G_{r}$, $1/G_{2}%
=1/g_{2}+1/2G_{r}$ with $g_{1}$ and $g_{2}$ being conductances of the R1%
$\vert$%
N1 and N2%
$\vert$%
R2 interfaces, respectively, $\alpha^{\prime}=(\alpha+\alpha_{se})(N_{x}%
+N_{y})/\left(  2\sqrt{N_{x}N_{y}-N_{xy}^{2}}\right)  $ (note that, when
important, bulk scattering can be easily included into the conductances
$g_{1}$, $g_{2}$ and $G_{r}$).\cite{Kovalev:prb02,Kovalev:prb06}

By solving Eqs. (\ref{LLGpump}-\ref{spin}) to the second order in a small rf
magnetic field the generated DC voltage can be expressed via the
susceptibilities Eqs. (\ref{suspumpx}) and (\ref{suspumpy}) as:%
\begin{equation}
eV=2\hbar\omega PG_{r}\left(  \frac{1}{g_{1}}-\frac{1}{g_{2}}\right)
\operatorname{Im}\left(  \chi_{\mathrm{xx}}\chi_{\mathrm{yx}}^{\ast}\right)
h_{x}^{2}/M_{s}^{2}, \label{DC}%
\end{equation}
where we introduced the effective polarization of the device $P=(1/G_{\uparrow
}-1/G_{\downarrow})/(1/G_{\uparrow}+1/G_{\downarrow}+4/g_{1}+4/g_{2})$. The
asymmetry of the interfaces R1%
$\vert$%
N1 and N2%
$\vert$%
R2 is seen to be crucial for the generation of a DC voltage in this geometry.
Note that the dependence of the DC voltage on the frequency has a Lorentzian
line shape as follows from Eqs. (\ref{suspumpy}-\ref{DC}). When $P>0$, the
positive sign of the voltage corresponds to the voltage applied to the
junction with higher conductance $g_{1(2)}$.

The first order correction in the mixing conductance $G_{i}$ leads to a slight
change of the height of the Lorentzian and to a shift of the resonance
frequency:%
\begin{equation}
\omega^{2}=\omega_{m}^{2}(1-{\frac{\gamma\hbar^{2}\left(  G_{1}^{2}+G_{2}%
^{2}\right)  }{4eM_{s}V_{m}}}\frac{G_{i}}{G_{r}^{2}})
\end{equation}
Note that the magnetovibrational coupling does not change the form Eq.
(\ref{DC}), when the proper susceptibilities Eqs. (\ref{MVR1}) are
substituted. The magnetovibrational coupling can thus be observed as a
splitting of the Lorentzian peak due to spin pumping in an asymmetric
N1$|$F$|$N2 structure. Generation of a DC voltage by magnetization precession
of a single ferromagnetic layer has been suggested in Ref.
\onlinecite{Wang:cm06}. However, those authors concentrate on the generation
of a DC voltage due to spin-flip scattering in the ferromagnet.

\section{Spin pumping and rectification of AC currents in F%
$\vert$%
N%
$\vert$%
F%
$\vert$%
N structures}

It follows from the previous appendix on N1$|$F$|$N2 structures that a
spin-coherent F1%
$\vert$%
N%
$\vert$%
F2%
$\vert$%
N structures in which the normal metals are identical and an extra
ferromagnetic layer F1 (with fixed magnetization) causes the asymmetry, should
generate a DC voltage as well. In such a structure, AC currents instead of the
rf magnetic field can generate spin transfer torques when the magnetizations
are non-collinear. Since the analytical expressions for arbitrary angles
between the magnetizations are complex, we concentrate here, without loss of
generality, on a 90 degree configuration.

We assume a constant AC current bias on the system, thus forcing the extra
charge current due to spin pumping to vanish. Following the derivation in
Appendix B, and disregarding the imaginary parts of mixing conductances for
both interfaces, we arrive at the following expressions for the linear
response functions:%
\begin{equation}
\chi_{\mathrm{xI}}(\omega)=(M_{y}/I)_{\omega}=\frac{\hbar\gamma}{2eV_{m}}%
\frac{\eta_{1}\left[  \gamma M_{s}(N_{y}+N_{xy}\alpha)-i\omega\alpha_{x}%
^{se}\right]  }{\omega^{2}(1+\alpha_{x}^{se}\alpha_{y}^{se})-\omega
_{\mathrm{m}}^{2}(1+\alpha^{2})+i\omega\Delta}, \label{susCURRx}%
\end{equation}
\begin{equation}
\chi_{\mathrm{yI}}(\omega)=(M_{x}/I)_{\omega}=\frac{\hbar\gamma}{2eV_{m}}%
\frac{\eta_{1}\left[  \gamma M_{s}(N_{xy}+N_{x}\alpha)-i\omega\right]
}{\omega^{2}(1+\alpha_{x}^{se}\alpha_{y}^{se})-\omega_{\mathrm{m}}%
^{2}(1+\alpha^{2})+i\omega\Delta}, \label{susCURRy}%
\end{equation}
where the anisotropic Gilbert dampings due to spin emission are $\alpha
_{x}^{se}=\gamma\hbar^{2}\left(  G_{1}+G_{2}\right)  /\left(  2eM_{s}V_{m}\right)
$, $\alpha_{y}^{se}=\gamma{\hbar^{2}\left(  G_{\uparrow\downarrow}%
+G_{2}\right)  /}\left(  2eM_{s}V_{m}\right)  $, $\Delta=\gamma M_{s}(N_{x}%
(\alpha+\alpha_{x}^{se})+N_{y}(\alpha+\alpha_{y}^{se})+\alpha(\alpha_{y}%
^{se}-\alpha_{x}^{se})N_{xy})$, $1/G_{1}=(1/g_{1\uparrow}+1/g_{1\downarrow
})/4+1/2G_{r}$, $1/G_{2}=1/g_{2}+1/2G_{r}$, $1/G_{\uparrow\downarrow}%
=1/2G_{r}+1/2g_{1r}$ with $g_{1\uparrow\downarrow}=g_{1r}+ig_{1i}$ (however
$g_{1i}$ is neglected) and $g_{1\uparrow(\downarrow)}$ being the mixing and
normal conductances of the R1%
$\vert$%
F interface, respectively.

The second order analysis provides us with an expression for the DC voltage.
After combining it with Eq. (\ref{u0}), we arrive at the full expression for
the DC voltage in terms of the susceptibilities Eqs. (\ref{susCURRx}%
,\ref{susCURRy}), consisting of separate contributions due to rectification
and spin pumping:
\begin{equation}
U_{0}=\frac{I_{0}^{2}}{2M_{s}}\frac{\partial R(\nu)}{\partial\nu
}\operatorname{Re}\chi_{\mathrm{yI}}+\frac{2\hbar\omega}{e}PG_{r}\left(
\frac{1}{\widetilde{g_{1}}}-\frac{1}{g_{2}}\right)  \operatorname{Im}\left(
\chi_{\mathrm{xI}}\chi_{\mathrm{yI}}\right)  I_{0}^{2}/M_{s}^{2}, \label{DC1}%
\end{equation}
where an effective conductance is $1/\widetilde{g_{1}}=\left(  3/2g_{1r}%
+(1/4g_{1\uparrow}+1/4g_{1\downarrow})R_{\uparrow\downarrow}/R_{1}\right)  /4$
and an effective polarization of the device is $P=(1/4G_{\uparrow
}-1/4G_{\downarrow})/(1/4G_{\uparrow}+1/4G_{\downarrow}+1/2g_{1r}+1/g_{2})$.
The dependence of the first term in Eq. (\ref{DC1}) on the frequency is in
general a combination of the Lorentzian with an antisymmetric Lorentzian, as
it is discussed in the Section IIIA. The second term becomes important when
$\alpha^{se}\gtrsim\alpha$ and ${\displaystyle \frac{1/\widetilde{g_{1}%
}-1/g_{2}}{1/g_{1\uparrow}-1/g_{1\downarrow}}}\gtrsim1$. When the damping due
to spin emission is small or the asymmetry of the tri-layer weak, the second
term in Eq. (\ref{DC1}) can be disregarded. Its dependence on the frequency
has a Lorentzian shape, see Eqs. (\ref{susCURRx},\ref{susCURRy}). In general,
the signs of the first and second term in Eq. (\ref{DC1}) can be opposite,
thus possible suppressing the symmetric-Lorentzian part of DC voltage.

The first order corrections in a small mixing conductances, $G_{i}$ and
$g_{1i}$ lead to a change of the height of the symmetric and antisymmetric
parts of Lorentzian and also to a shift of the resonance frequency, as it
arises from the expressions for the susceptibilities:%
\begin{equation}
\chi_{\mathrm{xI}}(\omega)=(M_{y}/I)_{\omega}=\frac{\hbar\gamma}{2eV_{m}}%
\frac{i\omega(\eta_{2}-\eta_{1}\alpha_{x}^{se})+\gamma M_{s}\Gamma_{x}}%
{\omega^{2}(1+\alpha_{x}^{se}\alpha_{y}^{se}+2\kappa)-\omega_{\mathrm{m}}%
^{2}(1+\alpha^{2})+i\omega\Delta}, \label{susCURRx1}%
\end{equation}
\begin{equation}
\chi_{\mathrm{yI}}(\omega)=(M_{x}/I)_{\omega}=\frac{\hbar\gamma}{2eV_{m}}%
\frac{-i\omega(\eta_{1}+\eta_{1}\kappa-\eta_{2}\alpha_{y}^{se})+\gamma
M_{s}\Gamma_{y}}{\omega^{2}(1+\alpha_{x}^{se}\alpha_{y}^{se}+2\kappa
)-\omega_{\mathrm{m}}^{2}(1+\alpha^{2})+i\omega\Delta}, \label{susCURRy1}%
\end{equation}
where $\kappa=\gamma{\displaystyle \frac{\hbar^{2}}{4eM_{s}V_{m}}}%
{\displaystyle (\frac{1}{R_{\uparrow\downarrow}R_{1}}+\frac{1}{R_{2}^{2}%
})\frac{G_{i}}{G_{r}^{2}}}$, $\Gamma_{x}=\eta_{1}(N_{y}+N_{xy}\alpha)+\eta
_{2}(N_{xy}-N_{y}\alpha)$ and $\Gamma_{y}=\eta_{1}(N_{xy}+N_{x}\alpha
)+\eta_{2}(N_{x}-N_{xy}\alpha)$.

Finally, we present our results in the presence of the magnetovibrational
coupling (\textit{e.g.} see Fig. \ref{rev2}) keeping only the dominant terms
in small Gilbert damping:%
\begin{equation}%
\begin{array}
[c]{ll}%
{\chi_{\mathrm{xI}}(\omega)} & {\displaystyle {={\displaystyle \frac{\hbar
\sin\theta\gamma}{2eV_{m}}}}}\\
& {\displaystyle {\times\frac{i\omega(\eta_{2}-\eta_{1}\alpha_{x}%
^{se}+g(1-N_{xy}/N_{x})F(\omega))+\gamma M_{s}\Gamma_{x}}{\omega^{2}%
(1+\kappa)-\omega_{\mathrm{m}}^{2}+i\omega\Delta+\Lambda F(\omega)}}}\\
& {\displaystyle \overset{g\rightarrow0}{\approx}{\frac{\hbar\sin\theta\gamma
}{2eV_{m}}\frac{\eta_{2}i\omega+\gamma M_{s}\Gamma_{x}}{\omega^{2}-\omega
_{\mathrm{m}}+i\omega\Delta+\Lambda F(\omega)},}}%
\end{array}
\label{MVRappX}%
\end{equation}
\begin{equation}%
\begin{array}
[c]{ll}%
{\chi_{\mathrm{yI}}(\omega)} & {\displaystyle {={\displaystyle \frac{\hbar
\sin\theta\gamma}{2eV_{m}}}}}\\
& {\displaystyle {\times\frac{-i\omega(\eta_{1}+\eta_{1}\kappa+\eta_{2}%
\alpha_{y}^{se})+\gamma M_{s}\Gamma_{y}\left[  1+g\left(  H_{z}/M_{s}N_{x}%
^{1}\right)  F(\omega)\right]  }{\omega^{2}(1+2\kappa)-\omega_{\mathrm{m}}%
^{2}+i\omega\Delta+\Lambda F(\omega)}}}\\
& {\displaystyle \overset{g\rightarrow0}{\approx}{\frac{\hbar\sin\theta\gamma
}{2eV_{m}}\frac{-\eta_{1}i\omega+\gamma M_{s}\Gamma_{y}}{\omega^{2}-\omega
_{\mathrm{m}}+i\omega\Delta+\Lambda F(\omega)}.}}%
\end{array}
\label{MVRappY}%
\end{equation}
These susceptibilities can be used in Eq. (\ref{DC1}) in order to calculate
the generated DC voltage. In the regime of resonant magnetovibrational
coupling, the second (spin emission) term in Eq. (\ref{DC1}) is comparable
with the first one only when $\alpha^{se}\sim\beta/\omega_{m}$. The second
term corresponds to two symmetric Lorentzian peaks split by $\operatorname{Re}%
\Lambda$.

Spin pumping therefore can play an important role when the enhanced interface
damping is comparable to the bulk Gilbert damping and, in the regime of
magnetovibrational coupling, the mechanical damping.

\end{document}